\documentclass[12pt]{article} 
\usepackage{hyperref}       
\usepackage{url}            
\usepackage{booktabs}       
\usepackage{amsfonts}       
\usepackage{microtype}      
\usepackage{newtxtext,newtxmath}
\usepackage{lipsum}
\usepackage[authoryear]{natbib}
\usepackage{color}         
\hypersetup{
colorlinks=true,
citecolor  =[HTML]{8B0000},
linkcolor  =[HTML]{8B0000},
urlcolor   =[HTML]{8B0000}
}
\usepackage{multirow}
\usepackage{makecell}
\usepackage{comment}
\usepackage{wrapfig}
\usepackage{setspace} 
\usepackage{amsmath}
\usepackage[inline]{enumitem}
\usepackage{arydshln}
\usepackage[most]{tcolorbox}
\usepackage{subcaption}
\usepackage{graphicx}
\usepackage{float}
\usepackage{needspace}
\usepackage{lipsum}
\usepackage{adjustbox}
\usepackage{framed}
\usepackage{xspace}
\usepackage{bbm}
\usepackage{diagbox}
\usepackage{caption}
\usepackage[most]{tcolorbox}
\definecolor{personacol}{HTML}{A6483E}   
\definecolor{situationcol}{HTML}{3A6B8A} 
\newtcolorbox{promptpart}[2]{%
enhanced, before skip=3pt, after skip=3pt,
colback=#1!5, colframe=#1!45, boxrule=0.5pt, arc=2pt,
left=7pt, right=7pt, top=4pt, bottom=4pt,
fonttitle=\footnotesize\bfseries,
coltitle=#1!70!black, colbacktitle=#1!32,
titlerule=0pt, title={#2}}
\newcommand{\orsep}{\vspace{1pt}{\centering\footnotesize\itshape\color{black!55}or\par}\vspace{1pt}}

\newtcolorbox{promptbox}{%
enhanced, colback=white, colframe=black!25, boxrule=0.4pt, arc=2pt,
left=6pt, right=6pt, top=5pt, bottom=5pt}
\newtcblisting{personaprompt}[1]{%
enhanced, breakable, listing only,
before={\par\Needspace{15\baselineskip}},
colback=personacol!5, colframe=personacol!45,
boxrule=0.5pt, arc=2pt,
left=7pt, right=7pt, top=5pt, bottom=5pt,
fonttitle=\small\bfseries,
coltitle=personacol!70!black, colbacktitle=personacol!32,
title={#1},
listing options={basicstyle=\small\rmfamily,breaklines=true,
breakatwhitespace=true,columns=fullflexible,keepspaces=true,
showstringspaces=false}}
\newtcblisting{gameprompt}[1]{%
enhanced, breakable, listing only,
before={\par\Needspace{15\baselineskip}},
colback=situationcol!5, colframe=situationcol!45,
boxrule=0.5pt, arc=2pt,
left=7pt, right=7pt, top=5pt, bottom=5pt,
fonttitle=\small\bfseries,
coltitle=situationcol!70!black, colbacktitle=situationcol!32,
title={#1},
listing options={basicstyle=\small\rmfamily,breaklines=true,
breakatwhitespace=true,columns=fullflexible,keepspaces=true,
showstringspaces=false}}
\newsavebox{\settingsbox}
\savebox{\settingsbox}{%
\begin{minipage}{0.46\textwidth}
{\footnotesize\itshape\color{black!60} Type responds to settings (one-at-a-time):\par}
\vspace{1pt}
\begin{promptpart}{situationcol}{Setting: Dictator Game}
\footnotesize You are splitting a pot of money between yourself and another\dots
\end{promptpart}
\orsep
\begin{promptpart}{situationcol}{Setting: Trust Game}
\footnotesize You are deciding how much money to send to another person\dots
\end{promptpart}
\orsep
\begin{promptpart}{situationcol}{Setting: Beauty Contest}
\footnotesize You are choosing a number to be closest to\dots
\end{promptpart}
\vspace{2pt}
{\centering\color{black!55}$\vdots$\par}
\end{minipage}}

\newcommand{\NumCountries}{35 }
\newcommand{\TotalSubjects}{78,657 } 
\newcommand{\TotalDecision}{119,147 } 
\newcommand{\IndivSubjects}{1,734 }
\newcommand{\IndivDecisions}{9,269 }
\newcommand{\PaperTitle}{Modeling Human Behavior with Type Vectors Using AI}

\title{\PaperTitle\thanks{
Corresponding author: Jackson. 
Affiliations: M.O.J. Stanford University, Santa Fe Institute, Monash University;  B.S.M.  MIT, MIT Initiative on the Digital Economy;  Y.X. University of Michigan, University of Chicago, Vanderbilt University; W.Y. Moblab; Q.M. University of Michigan.
Competing interest statement: W.Y. is the Chief Executive
Officer (CEO) of MobLab. M.O.J. is the
Chief Scientific Advisor of MobLab and Q.M. is a Scientific
Advisor to MobLab, positions with no compensation but
with ownership stakes. 
B.S.M. is an advisor to and has an ownership stake in Expected Parrot.
Y.X. has no competing interests.
In preparing this paper, the authors utilized generative AI models as tools in their main analysis and to copyedit.
}}

\author{
Matthew O. Jackson\footnote{Indicates co-first authors, ordered alphabetically.}  \and
Benjamin S. Manning\footnotemark[\value{footnote}]  \and 
Yutong Xie\footnotemark[\value{footnote}]  \and
Walter Yuan \and
Qiaozhu Mei 
}

\begin{document}
\maketitle

\begin{abstract}
\noindent We introduce a general, easy-to-implement AI-based modeling technique for analyzing human behavior.
A key feature of this approach, which contrasts with existing modeling techniques, is that it combines the flexibility and interpretability of natural language with a mathematical structure that can be fitted to data and easily analyzed.
We assign a large language model a vector of trait intensities---a type vector---and then ask it to choose actions across settings in which we observe human choices.
For instance, the type vector $(2,4)$ could correspond to ``You are a player characterized by the following profile: Altruism: 2 out of 5, Risk Aversion: 4 out of 5,'' after which it is asked to make choices.
We can then vary the traits (e.g., Altruism, Fairness, Trust, $\dots$) and values (e.g., 1--5) to minimize distance to human choices.
We illustrate the method by applying it to model \TotalDecision decisions made by \TotalSubjects subjects from more than \NumCountries countries across 10 classic economic game roles.
We find that human behavior can be closely matched using three dimensions: Risk Aversion, Strategic Sophistication, and Trust.
The type vectors needed to fit individuals across games cluster into fewer than a dozen groups, with substantial variation in fit across subjects.
Moreover, the individual type vectors can predict behavior in held-out games with different rules and available actions.  
More broadly, this new modeling method is highly generalizable and interpretable: we can input any vector of traits and use them to model behavior across any setting.
\end{abstract}

\thispagestyle{empty}

\setcounter{page}{0}
\newpage \clearpage

\singlespacing

\section{Introduction}

Humans are heterogeneous in preferences and behaviors, and are motivated beyond the selfish attitudes of classic Homo-Economicus.
Not only do humans vary in terms of preferences, attitudes towards risk, and strategic sophistication \citep{holt2002risk,nagel1995unraveling,camerer2004cognitive}, but they are also swayed to varying degrees by social aspects such as altruism, fairness, and trust \citep{charness2002understanding,andreoni2002giving,berg1995trust}.  
A plethora of theories from the social sciences explain various behaviors, especially those that deviate from fully sophisticated and self-interested individuals \citep{camerer2003behavioral,kahneman1979prospect,laibson1997golden}. 
Such theories are typically designed to explain a particular feature of human behavior, like the role of fairness in bargaining and how it justifies rejection of positive offers in an ultimatum game \citep{fehr1999theory,bolton2000erc}, or how conscientiousness (one of psychology's ``Big 5'') relates to educational attainment \citep{becker2012relationship}.

This menagerie of setting-specific theories raises fundamental questions: is human behavior high-dimensional, or can just a few characteristics jointly account for choices across environments?
And whatever this space may be, do people span the space broadly or cluster into a few recurring types?
How heterogeneous are people in terms of how well they can be fit across environments?
Our main contribution is to introduce a tractable, portable, and easy-to-implement new modeling technique that uses generative AI to fit, estimate, and analyze human behavior across diverse settings.
We then illustrate it by answering the questions above for a set of classic economic game roles.   

A growing literature shows that Large Language Models (LLMs) can be prompted to imitate human behavior with high fidelity across a wide range of settings \citep{horton2023large,xie2025using,jackson2025ai,akata2025playing,manning2025general,ashokkumar2026large}.
Some of this work, along with other research \citep{serapiogarcia2025psychometric,wang2025evaluating,huang2026designing,sakai2026effects}, has also shown that ``AI simulations'' can be systematically tuned by assigning numeric operationalizations of concepts within prompts.
We use this tunable format, but rather than as a method to simulate human behavior with the goal of prediction, we instead use it to model and better understand human behavior.
A key feature of our approach is that it combines the flexibility and interpretability of natural language, which has been taken advantage of by the simulations literature, with a simple mathematical structure that serves as a model that can be fitted to data and analyzed for insights.

Our modeling method structures an input as a set of traits or dimensions represented as a vector, which can have varying intensities.
We then prompt an LLM with a natural language description of this vector together with a scenario in which it is asked to choose a behavior.
For example, if we are interested in how Altruism drives behavior in certain settings, we can prompt an LLM with instructions saying that it has the trait ``Altruism: level $x_1$ out of 5.'' 
We can then ask it what actions it would choose in a series of different scenarios---like a dictator or trust game.
As $x_1$ varies, the LLM generates different behaviors in a given scenario or set of scenarios.
We can similarly use Risk Aversion, or combine multiple characteristic traits by prompting the LLM that it has the profile ``Altruism: $x_1$ out of 5, Risk Aversion: $x_2$ out of 5,'' and so forth. 
The resulting profile $(x_1,x_2)$ represents a type as a vector of intensities for the characteristics. 
The LLM then maps this type vector into behavior in whichever settings we input.
Thus, we can then estimate which ``type vectors'' best fit human behaviors across any settings described in natural language.
These fitted vectors provide a simple mathematical representation of traits described in the prompt, which are far simpler than the underlying embeddings or the LLM's internal representations.
This structure lets us examine how many traits are needed to fit behavior, and how individual types cluster and relate to one another, among other questions.

In the empirical portion of this paper, we illustrate our method by building type vectors using five traits that figure prominently in behavioral economics: Altruism, Fairness, Risk Aversion, Strategic Sophistication, and Trust.
We apply the method to ten classic economic game roles.
These roles exhibit substantial heterogeneity in human behavior and span various prominent theories. 
As we vary the profiles $(x_1,x_2,...)$, we measure how closely the LLM's choices match human behavior across the games. 
The profiles are varied both in the numeric values in the entries and in the set of traits included---that is, the content and dimensionality of the vectors (up to five dimensions).
In this way, our method enables us to estimate an upper bound on the number of dimensions needed to closely approximate human behavior across the game-roles we study.
For example, if varying a simple three-dimensional prompt---like one where the traits are Altruism, Fairness, and Trust, each varying on a scale of 1-5---can match individuals well across the ten settings, then, regardless of the content of those prompts, it means that individual humans can be approximated and predicted across a variety of situations, with just a simple three-dimensional type. 
Part of our contribution is to provide evidence that such an upper bound is credible.

Our analysis proceeds on a dataset of \TotalDecision decisions made by \TotalSubjects participants across more than \NumCountries countries who played various combinations of the ten game roles.
First, we analyze how well we can match the distribution of human behaviors within each game role by varying the prompts with our five primary dimensions.
The distribution fits across the games are generally extremely tight.
We also find intuitive patterns between which prompts are most valuable in matching behaviors and how that varies across games.
For instance, as one should expect, Risk Aversion plays a key role in matching behavior in both the bomb game (used to assess risk preferences) and the trust game (in which a player's payoff depends on the reciprocity of their partner), among others.
In fact, Risk Aversion alone does fairly well at matching behaviors in several games. 
Altruism also does well at matching behavior when the other player's payoff depends on a given subject's behavior.
Ultimately, we find that human behavior can be closely matched using just three dimensions: Risk Aversion, Strategic Sophistication, and Trust.

Beyond matching distributions of humans playing each game role, we then match individual humans across game roles.
The challenge is not just to match the marginal distribution of behavior in each game, but to match the joint distribution across games. 
We do so on a subset of \IndivDecisions decisions from \IndivSubjects subjects who played at least five game roles.
This presents an additional challenge: a subject's behaviors across multiple games must be matched with one type vector.

We find substantial heterogeneity in the number of dimensions needed to approximate individual subjects.
Some can be well-matched with just one or two dimensions, while  others require more.
Even though there are more than a thousand potential individual type vectors, the types that best fit the humans break into fewer than a dozen distinct, tight clusters.
For example, some participants are both highly strategic and risk-seeking, others are highly risk-averse and fair, but we do not observe clusters that are both highly risk-averse and highly strategic.

We also perform an out-of-distribution analysis to understand whether individual types generalize to new settings.
We withhold a subject's play in a given game, assign them a best-fitting economic trait type vector based on their play in the other games, and then see how closely the play induced by that type on the held-out game matches the subject's actual play. 
We compare these predictions to a benchmark that has full in-distribution information, building Machine Learning (ML) models that predict a subject's behavior in one game role from all their plays across all other game roles.
Compared to this benchmark, our out-of-distribution type assignments match almost all subjects well.

A key feature of our modeling method is its flexibility.
Researchers can construct type vectors based on any social science theory, and then use the LLM to translate those types into behaviors across different settings.
To illustrate this, we compare the performance of our five economic dimensions to that of psychology's Big 5 personality traits, and also to various atheoretical traits.
The economic characteristics perform best in our analysis, but the broader result does not depend on interpreting the characteristics literally.
A different or better-chosen set of characteristics might perform even better or use fewer dimensions.
Thus, the curve obtained from our estimates provides an upper bound for the settings we study.

From complexity theory, we know that combinations of a few dimensions can produce complex patterns \citep{krakauer2024complex}, and so our results do not suggest that human behavior is simple. 
Rather, it can be well-mimicked with a low-dimensional model, at least across the ten game roles we study.
Given that the behavioral literature has largely progressed with theories of specific behaviors in particular scenarios, this provides further empirical motivation for building meta-theories with only a few moving parts. 
Our contribution is both to provide evidence for this possibility and to introduce a general method for developing and analyzing representations of behavior across settings.

\paragraph{Related Research}

Our work connects to a literature using ML to evaluate and develop models of human behavior.
This research uses ML to assess economic theories \citep{fudenberg2022measuring,andrews2025transfer,ellis2026predicting} and
identify regularities to inform new theories \citep{fudenberg2019predicting,peterson2021using,ludwig2024machine,zhu2025capturing,mullainathan2025anomalies,liang2026using}.
As mentioned above, unlike approaches that require model adjustments for particular settings, our technique lets researchers to apply a model to settings without altering the model and easily evaluate competing theories.

A related strand uses LLMs to construct AI simulations or ``digital twins'' \citep{argyle2023out,aher2023using,binz2023using,park2023,park2024generative,kim2023ai,lippert2024can,yeykelis2024using,kozlowski2024silico,manning2024automated,binz2025foundation,qian2026strategicAI,gao2026llm,abdel2026life}. 
That research studies whether LLMs can be prompted to reproduce the behavior of human populations or particular individuals, as well as the conditions under which such simulations are reliable \citep{dillion2023can,chen2023emergence,ross2024llm,abdurahman2024perils,anthis2025position,kozlowski_evans_2025,hullman2026involve,peng2026funhouse}.
Similar to that literature, our method depends on AI's steerability and ability to emulate patterns of human behavior, which is setting-dependent but rapidly improving. 
But in contrast to the AI simulations literature, we use LLMs as a tool for modeling, measuring, and analyzing the structure and underlying complexity of human behavior.
In this way, our work builds on \citet{xie2025using}, which used much longer prompts to study human motivations.
Here, we reduce those prompts to vectors of single-word characteristics whose intensities can be tuned, and use them to study how many dimensions are needed to approximate human behavior.

Our application relates to previous work on the dimensionality of human behavior.
\citet{fudenberg2006advancing} argues that behavioral economics should produce more unified explanations that apply across a wider range of phenomena. 
Consistent with this call, a growing body of theoretical work develops and tests more parsimonious foundations of human behavior through mechanisms like limited
attention \citep{gabaix2014sparsity}, imperfect memory \citep{bordalo2020memory}, and imperfect cognition \citep{woodford2020imprecision,enke2023cognitive,enke2024attenuation}.
More related to our application, a similarly motivated strand of research searches for empirical evidence of simple shared structure across behaviors.
\citet{dean2019empirical} study relationships among eleven economic behaviors to assess whether they can be explained by a parsimonious model of choice. 
\citet{bruhin2019many} show that social behavior can be characterized by a small number of stable types that predict behavior across games.
Most related to our work in terms of motivation are \cite{chapman2023econographics}, who examined correlations across 21 questions and, via a principal component analysis, found that the first 6 components capture much of the variation.
Similarly, \citet{stango2023taxonomy} measure a broad set of behavioral tendencies and summarize them using a small number of common factors.
Our results echo the findings that human behavior across settings has a low-dimensional structure, but we arrive at this conclusion using a very different methodology, which thus adds a layer of robustness to the finding.
Moreover, as we show, our method has additional benefits and can be particularly useful in out-of-distribution analyses.

A related thread of research in psychology studies the many ways in which people differ using a few dimensions; for example, psychology's Big 5 personality traits \citep{costa1992revised}. 
Relatedly, \citet{becker2012relationship} and \citet{jagelka2024economists} study the relationship between economic preferences and the Big 5. 
Our comparison between economically derived dimensions and the Big 5 is similarly motivated.

We emphasize that our technique is different from and complementary to such previous dimension analyses, and provides a different insight from methods like principal component analysis, factor analysis, LASSO, and other statistical and ML methods.
While such approaches provide taxonomies of observed behaviors, we construct types that can generate behavior across (arbitrary) settings---anything that can be described in natural language. 
Thus, our technique is completely portable and does not need to be tailored to the details of any out-of-sample setting.
As \cite{manning2025general} highlight, because both the type and the setting are described in natural language, the same representation can be applied without redesign to environments with different action spaces, rules, payoffs, or player structures.

\section{Method: Modeling Human Behavior Using AI and Type Vectors }
\label{sec:prompt-method}

In this section, we introduce the method, compare and contrast it with other modeling techniques, and discuss the new kinds of questions it lets us explore.

\subsection{Type Vectors as Inputs to AI}

Our approach represents ``types'' as vectors over a set of behavioral traits.
A researcher begins by specifying a set of characteristics that they hypothesize will produce variations in behavior as those characteristics are varied.
These characteristics can be drawn from economics, psychology, or any other theories of human behavior. Also, the type vector can include things that are not only innate traits and preferences, but also things like beliefs about the world in the standard sense of  \citet{harsanyi1967games}.

Specifically, a type vector assigns an intensity to each characteristic.
An LLM is given a description of the type vector in words, which, when coupled with descriptions of settings (e.g., games, surveys, etc.), produces choices of behaviors for this type across those settings.
By examining the behaviors induced as we change the number of characteristics included in the type vector, we can ask how many dimensions are needed to closely approximate the behavior of human populations, as well as individual people, across any given collection of settings of interest.
This also allows us to represent individuals with potentially meaningful types.

\begin{figure}[h]
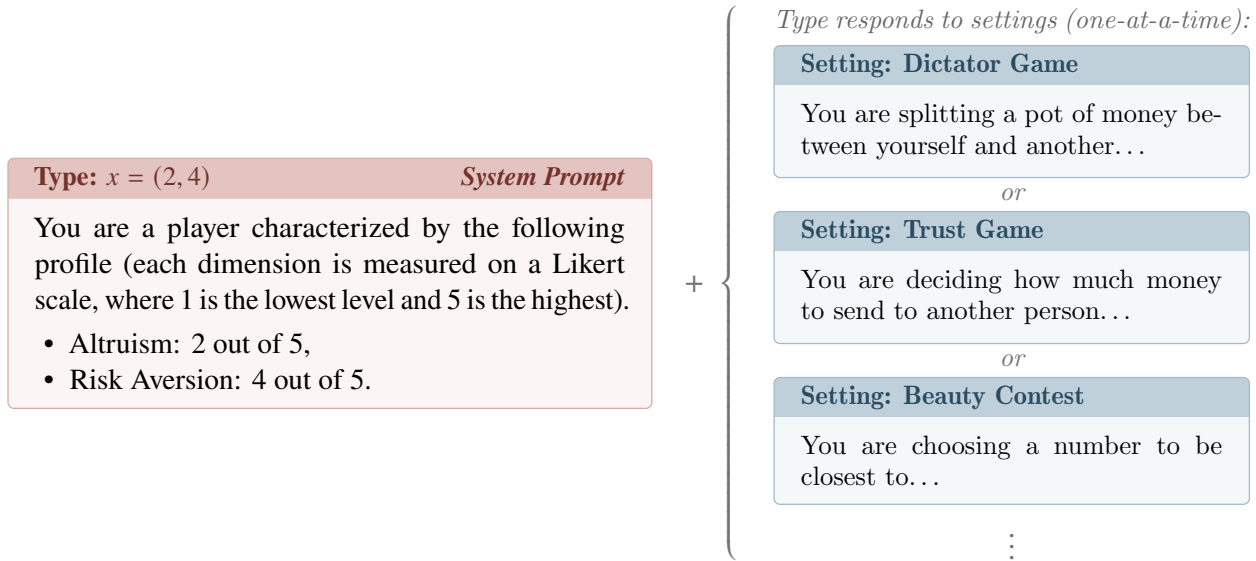

\centering
\begin{minipage}{\textwidth}
\centering
\begin{minipage}[c]{0.48\linewidth}
\begin{promptpart}{personacol}{Type: $x=(2,4)$\hfill \emph{System Prompt}}
\small You are a player characterized by the following profile
(each dimension is measured on a Likert scale, where 1 is the lowest level and 5 is the highest).
\begin{itemize}[leftmargin=1.2em,nosep,topsep=3pt]
\item Altruism: 2 out of 5,
\item Risk Aversion: 4 out of 5.
\end{itemize}
\end{promptpart}
\end{minipage}%
\begin{minipage}[c]{0.09\linewidth}
\centering\color{black!55}%
$+\;\left\{\rule[-\dp\settingsbox]{0pt}{\ht\settingsbox}\right.$
\end{minipage}%
\begin{minipage}[c]{0.5\linewidth}
\usebox{\settingsbox}
\end{minipage}
\end{minipage}
\caption{A type vector becomes the system portion of a prompt, which is then paired with the instructions for one setting at a time.
For example, the two-dimensional type
$x=(2,4)$ over $(\text{Altruism}, \text{Risk Aversion})$ can be paired with any setting that can be described in words, generating a separate response in each.}
\label{fig:method_schematic}
\end{figure}

To make this concrete, suppose that we want to match the behavior of a human subject by using Altruism and Risk Aversion. 
We assign each characteristic a level from 1 to 5, so that the vector $(2,4)$ describes a person who is a 2 out of 5 in Altruism and a 4 out of 5 in Risk Aversion. 
Figure \ref{fig:method_schematic} shows how this vector becomes a prompt, which can be paired with various descriptions of settings and questions about choices of behaviors to be made.
We supply the vector-based prompt, the setting, and a question about a choice, and the LLM responds with a behavior.
Thus, for each type vector and setting, this method generates a prediction.
Because such prompts operate entirely in natural language, one can apply any description of a type and setting and generate behaviors.
This could include economic games, the focus of our paper, or other scenarios.

More generally, suppose that a researcher begins with $D$ candidate characteristics and selects $k\leq D$ of them for a particular representation. 
Each selected characteristic is assigned a level on an ordered scale, and the resulting $k$-dimensional type is a vector $
x=(x_1,\ldots,x_k).$
Each characteristic is measured on an $L$-point Likert scale.
With $L$ possible levels, a fixed set of $k$ characteristics defines $L^k$ possible type vectors. 
In our implementation, $L=5$, so a $k$-dimensional representation contains $5^k$ possible types.
In addition, we can build lower-dimensional vectors by using only a subset of the characteristics with a smaller $k$. 
By varying the number of characteristics $k$, we can measure the number of dimensions needed to match human behavior for a particular set of trait characteristics.

\subsection{A Sixth Class of Models}

It is useful to situate this technique in the landscape of quantitative models used to analyze human behavior.  
A rough taxonomy consists of 
(i) statistical/econometric models, (ii) machine learning models, (iii) economic theory and behavioral models, (iv) agent-based models, (v) dynamic and evolutionary system models.
Of course, such a rough categorization cannot do justice to the full catalog of models, and these categories overlap, but it is nonetheless useful. 
In particular, each of these has a mathematical structure to the input which can be varied to fit data.
We then learn from seeing which math structure matches whichever data we wish to analyze.
One can then also use a model to do comparative statics and policy analysis.   

What we are introducing here is a sixth class of models, based on using AI, that has these above features, but also has three key advantages over previous modeling techniques. 
The first is that we can apply the same model of underlying human traits to any arbitrary setting  (which is not true of the other five model classes).
That is, we can input the same vector of traits and then record AI's behavior as we ask it to behave in any setting.
The second is that the traits are easy to introduce in natural language, so that any theory which can be expressed in levels of traits can be easily tested.
The third is that one can easily contrast and compare different theories across domains. 
Our technique thus complements standard modeling techniques.

The main limitation of our technique is that the AI portion of the model is a black box; at least if one uses an off-the-shelf LLM as we do here.
This is also true of many models in the other classes: for instance, a by-product of complexity theory is often that simple models can exhibit complex patterns in ways that are not easily grasped---e.g., chaos theory, fractals, and so on.  
One therefore has to be careful in interpreting the fits that emerge.
A good fit does not by itself establish that a model captures people's underlying motivations and reasoning.
What we can do is see what sort of structure fits which sort of data, without concluding that the prompted traits have a particular meaning.

With this in mind, the kinds of questions that we can analyze are how many traits are needed to match behavior across settings, how clustered humans are in terms of the type vectors needed to match them, how heterogeneous humans are in terms of ability to be fit, and whether some sets of traits fit behavior better than others.
These questions do not require direct interpretations of the traits.
Nonetheless, as we see in our application, there turn out to be intuitive relationships between how the traits work across games that nicely match the theories underlying those traits. 
This suggests that AI type vectors offer a promising approach for directly comparing theories based on how well vectors built based on corresponding traits fit human data.

\section{Data, Game Roles, and Candidate Dimensions}
\label{sec:game-data}

In this section, we introduce the game roles, the human dataset, and the candidate dimensions we study.
We also discuss the interpretation of the type space.

\subsection{Data and Game Roles}

Our data include human responses from eight classic games from the behavioral economics literature: a Bomb Risk game, a Commons Game, a Cournot Game, a Dictator Game,  an Investment Game, Keynes' Beauty Contest Game, a Public Goods Game, and an Ultimatum Game.
For two of the games---the Ultimatum and Investment games---we examine behavior in two different roles.
These are the Proposer and Responder for the Ultimatum Game, and the Investor and Banker for the Investment Game.
This gives us ten distinct game roles in which to analyze behavior. 
For each of the ten game roles, we have human data from the MobLab Classroom economics experiment platform, which consists of \TotalSubjects subjects from more than \NumCountries countries, spanning multiple years.\footnote{These data are largely from classrooms, and many are non-incentivized.
Nonetheless \cite{lin2020evidence} find that behavioral regularities in MobLab data are broadly consistent with laboratory experiments and robust across available incentive structures.
Regardless, our data are used to illustrate our method and to show that the rich heterogeneity in the data can be generated, quantified, and analyzed using a low-dimensional model.
None of that is dependent on the incentives in the data collection.}
The full instructions for each game are provided in Appendix~\ref{app:game-instructions}, and additional details about the human-playing data appear in Appendix~\ref{app:human-data}.

The top row of Figure \ref{fig:distributions} shows the distributions of human play across the ten different game roles.
We observe substantial heterogeneity in human play as well as spikes in the distributions.
The distributions include, but differ from, both the Nash equilibrium action and the play that maximizes the total sum of the players' payoffs.

\subsection{Candidate Dimensions}

The five characteristics that we focus on in this paper are: Altruism (A), Fairness (F), Risk Aversion (R), Strategic Sophistication (S), and Trust (T).
These were derived from a list of traits that figured prominently in \citet{xie2025using} and are well-studied in the behavioral economics literature.
In Section \ref{sec:alt-dims}, we also consider what happens when alternative characteristics are used instead, like placebos and psychology's Big-5 personality traits.
Our prompts follow the exact format of Figure~\ref{fig:method_schematic}.
For each type vector, we provide GPT-4.1 with a system prompt based on that type vector, followed by a user prompt based on the game instructions used to generate the human subject data.
Appendix~\ref{app:prompts} gives the exact prompts along with the model versions, collection dates, inference settings, response counts, and extraction procedure. 

\begin{figure}[h]
\centering
\includegraphics[width=\linewidth]{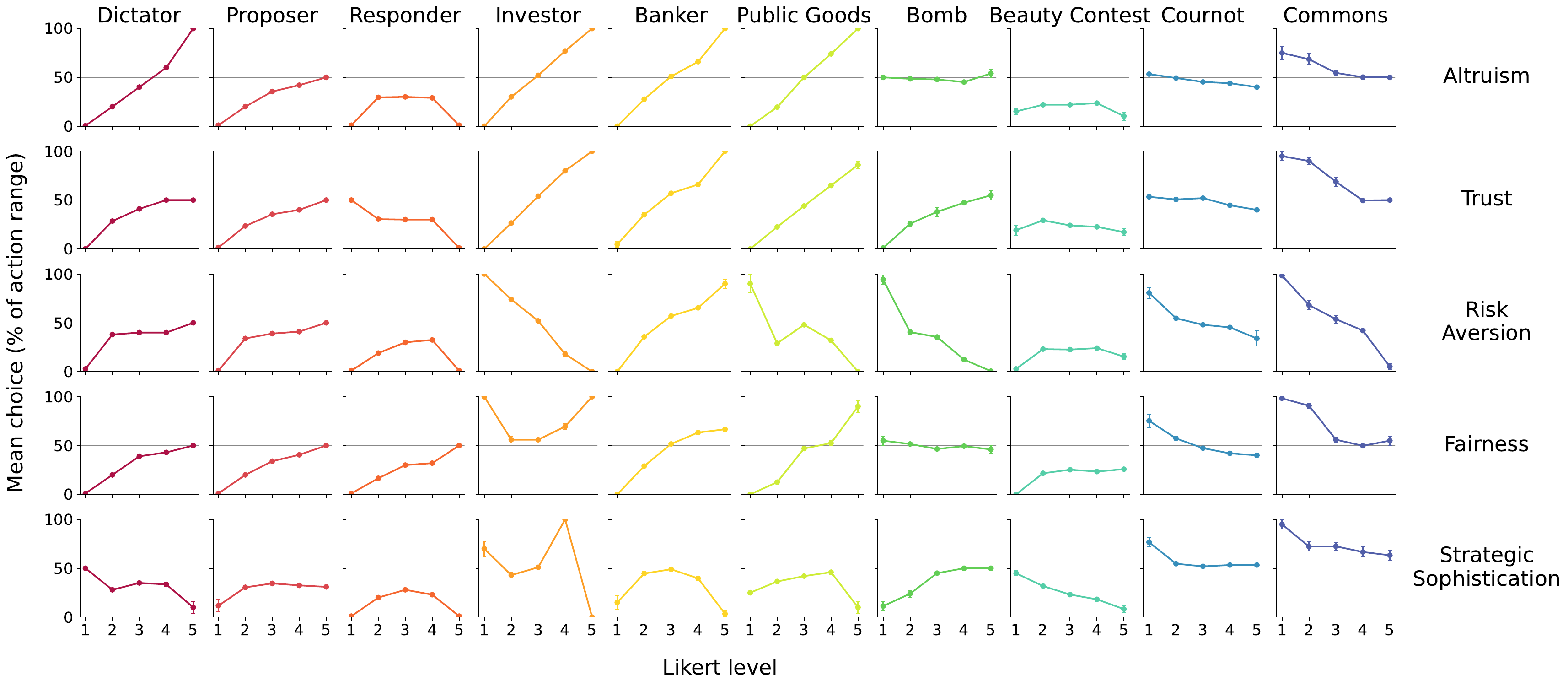}
\caption{For each game role, we show mean behavior as the Likert level of each behavioral characteristic dimension is varied in isolation (one-dimensional types).
Choices, the $y$-axis, are expressed as a percentage of each game's action range.
The horizontal black line provides a reference to help visualize the variation across scales.}
\label{fig:dimension_behavior_L5}
\end{figure}

Varying the 5 Likert levels of these five dimensions generates a collection of $L^k = 5^5 = 3,125$ behavioral type vectors.
Lower-dimensional type spaces ($k \in \{1,2,3,4\}$) are constructed by including only a subset of the characteristics. 
For each game role, the LLM then maps a type vector and the game instructions into a choice in the game. 
Figure \ref{fig:dimension_behavior_L5} illustrates how each characteristic affects behavior when varied as a single dimension type prompt in isolation.
That is, we prompt the LLM with only that characteristic and elicit its choice in each game for each of the five Likert values.
To make responses comparable across games, we normalize the choices as a percentage of each game's action range.

Effects differ systematically across games.
In some games, the LLM's choice varies in response to each of the five dimensions, while in other games the choices vary in response to only some of them. 
In general, choices vary in expected ways.
For example, Altruism has little effect in the Bomb, Beauty Contest, and Cournot games, but substantially affects behavior in the Dictator, Banker, and Public Goods games.
Strategic Sophistication, in contrast, does not impact choices for the proposer and responder roles, but generates monotonically decreasing choices for the Beauty Contest.

Our main analysis uses GPT-4.1, but nothing in the method is specific to this model.
Appendix~\ref{app:models} repeats the one-dimensional exercise using GPT-5.6 Terra, GPT-5.6 Luna, Claude Sonnet 5, and DeepSeek V4 Pro.
The characteristic-by-game relationships remain intuitive across models.
For example, Altruism consistently affects giving and investment, while Strategic Sophistication consistently lowers choices in the Beauty Contest.
More generally, each alternative model's 50 trait-by-game correlations are highly correlated with the corresponding GPT-4.1 correlations, with $r$ ranging from 0.83 to 0.88 (see Figure~\ref{fig:lfive-correlation}).

\subsection{The Dimensionality of the Type Space}
\label{sec:type-space}

Before moving to our empirical analysis, we briefly discuss how to interpret the types our method produces.
When we use a $k$-dimensional vector with five potential levels on each characteristic, this results in a Type space of $5^k$.
The human data lives in a space that has $100^{10}$ potential action profiles (10 games with 100 percentage-point levels in the action space).
With $k=3$, for instance, $5^3=125$ is much lower-dimensional than $100^{10}$, which alleviates concerns of simple overfitting.

Nonetheless, $125$ types is a lot of potential types.
If the LLM arbitrarily interprets types, a three-dimensional type vector could index 125 unrelated patterns of behavior.
Nearby points in the type space would not necessarily generate similar behavior, allowing the 125 types to approximate substantial human heterogeneity without forming a coherent low-dimensional space.
Because LLMs are highly nonlinear, counting prompted dimensions alone does not rule out this possibility.\footnote{See \citet{vafa2024world} for examples of LLMs with structured implied world models.}

This concern is allayed by the patterns we see in Figure \ref{fig:dimension_behavior_L5}.
First, the patterns are fairly continuous: changing the prompted intensity generally does not produce abrupt changes in behavior.
Second, the relationships are typically either flat or approximately monotonic (if not linear).
Some exhibit a reversal, but generally no more than one. 
In Appendix~\ref{app:bivariate-effects} we extend this exercise to two-dimensional type vectors by varying one characteristic while holding a second characteristic fixed at each of its five levels.
Even in this larger type space, the relationships generally remain smooth, often roughly linear, and either flat or approximately monotonic.
These patterns would not lead to arbitrary placement of points throughout a space, but indicate that nearby type vectors generally produce similar behavior.
 
\section{Results}

\subsection{Matching Distributions of Human Play} \label{sec:match_dist}

Our first analysis studies how well different combinations and numbers of dimensions generate the range of human behavior observed within each game.
For each number of dimensions from one through five, we consider every possible subset of that size from our five economic characteristics.
For every subset and game, we choose mixture weights over the resulting types to minimize the normalized Wasserstein distance between the AI and human distributions.
We fit the mixture weights separately for each game.

\begin{figure}[h]
\centering
\includegraphics[width=\linewidth]{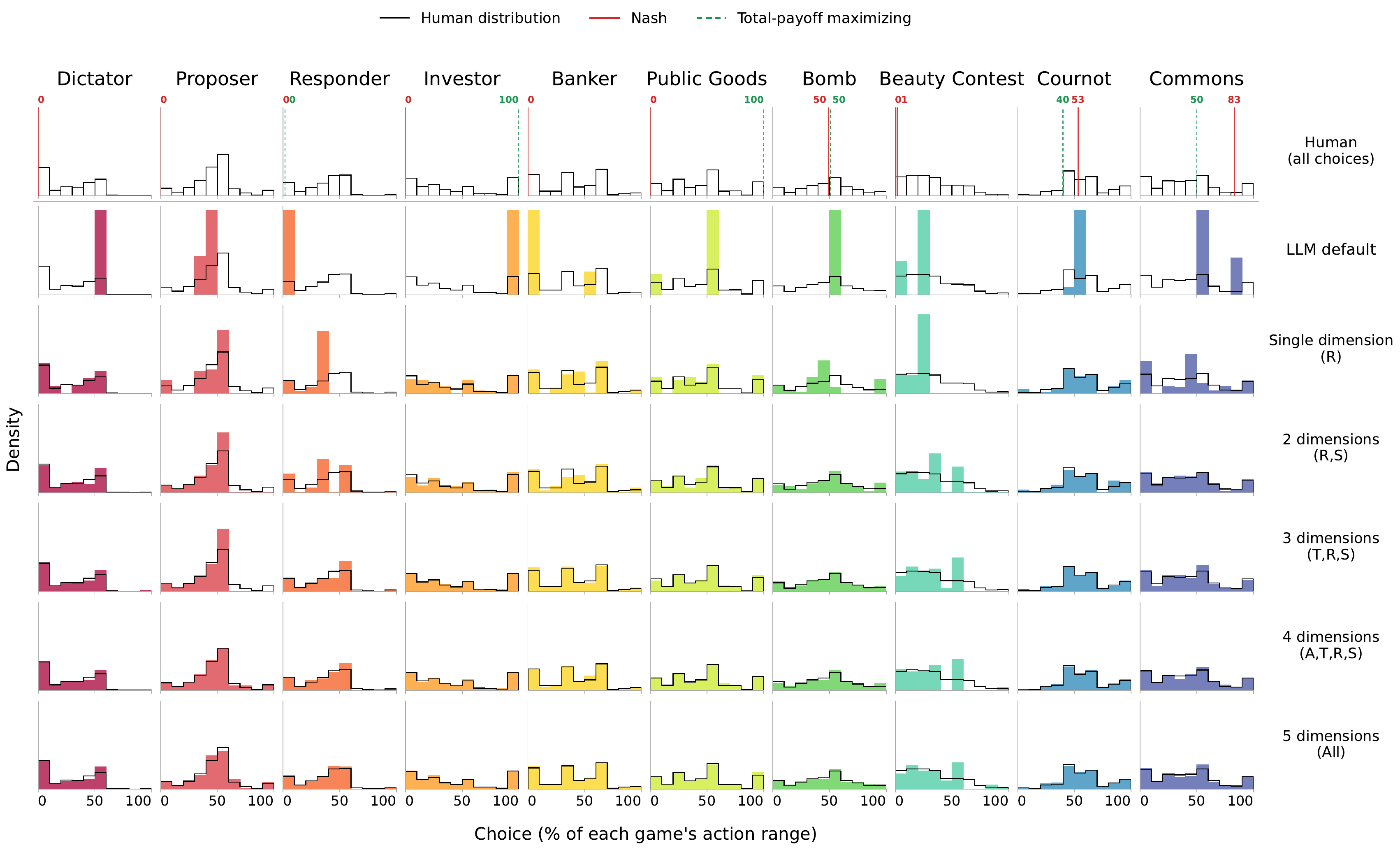}
\caption{Distributions of human and AI play across the ten game roles.
Choices are expressed as a percentage of each game's action range.
The top row shows the distribution of human choices, and the black outline repeats the human distribution in each subsequent panel.
The solid red and dashed green lines in the first row indicate the Nash and total-payoff-maximizing actions, respectively.
The second row shows choices under the default system prompt without specifying a type.
The remaining rows show 1,000 choices drawn from the best-fitting mixture of type-vector prompts using one through five dimensions.
For each number of dimensions, the type vector on the right is the combination with the lowest mean normalized Wasserstein distance across all ten game roles and is used in every column.
For example, the two-dimensional row uses Risk Aversion and Strategic Sophistication $(R,S)$, the pair with the lowest average normalized distance across games.}
\label{fig:distributions}
\end{figure}

For each number of dimensions, we identify the single subset of characteristics with the lowest average normalized Wasserstein distance across the ten game roles.
Figure~\ref{fig:distributions} shows the distributions generated by these best subsets, along with the empirical distribution of human play and the distribution generated by the default prompt, which contains no type vector.

Moving down the rows, we see that the default prompt generally produces choices concentrated on one or two points and fails to reproduce the heterogeneity in human play.
Even with one-dimensional types, however, the LLM distributions are substantially closer to the human distributions.
With three dimensions, the generated distributions largely overlap with the human distributions in most games, and adding fourth and fifth dimensions produces smaller additional improvements.

\begin{figure}[h]
\centering
\includegraphics[width=\linewidth]{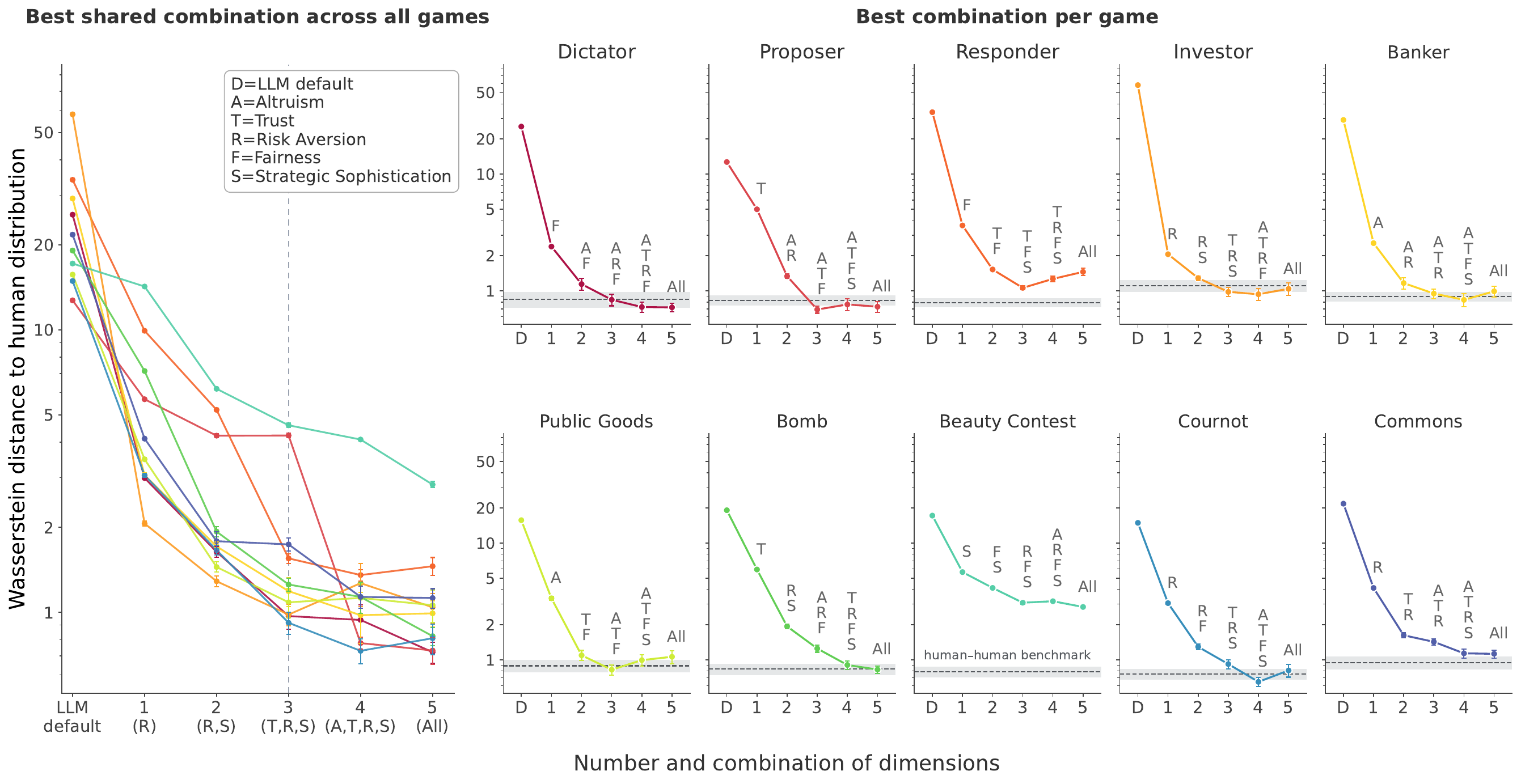}
\caption{
The left plot shows the Wasserstein distance between the responses of the various types and human play, averaged across game roles, by number of prompt dimensions, while the right plots show it for each game individually.
The dashed line and band in each game-role plot give the benchmark distance between 1,000 random draws of humans and the full human data ($\pm1$ s.e.).
For each game role and each $k$, mixture weights over the $5^K$ type vectors are fit to minimize the distance to the human distribution.
Each point is the distance between 1,000 draws from the fitted mixture and the full human data, normalized by the game's action range, averaged over 10 runs ($\pm1$ s.e.).
The point labeled $D$ is generated by the default prompt without a type vector.
In the left plot, each dimension $k$ uses the trait combination with the lowest mean distance across all ten game roles.
In the right plots, each game role uses its own best combination at each $k$, annotated beside each point.}
\label{fig:wd-progression}
\end{figure}

The left panel of Figure~\ref{fig:wd-progression} quantifies these patterns.
It plots the normalized Wasserstein distance between the elicited distribution and the human distribution for each game.
The first point on the $x$-axis shows the fit under the default prompt, and moving to the right shows the fit with increasing numbers of dimensions using the combinations listed in the right-hand margin of Figure~\ref{fig:distributions}.

The patterns in the left panel are consistent with the observations from Figure~\ref{fig:distributions}.
The largest improvements occur within the first three dimensions.
Moving from the default prompt to Risk Aversion alone substantially reduces the distance for every game.
Adding Strategic Sophistication and then Trust produces further improvements.
The fourth and fifth dimensions provide much smaller additional gains for all game roles other than the Proposer.
In sum, Risk Aversion, Strategic Sophistication, and Trust reproduce much of the distributional heterogeneity across games.

In Figure \ref{fig:wd-progression}, the ten smaller plots on the right allow each game to use its own best combination of characteristics.
The letters above each point identify this combination.
For example, the $F$ above the one-dimensional point in the Dictator panel indicates that Fairness provides the best one-dimensional fit for that game.
The dashed line and band in each smaller plot show the distance between random samples of 1,000 human choices and the full human distribution.
They therefore provide a benchmark for the distance produced by sampling error alone.

The best single dimensions are often intuitive: Fairness fits the Dictator and Responder roles best, Risk Aversion for the Investor, and Strategic Sophistication for the Beauty Contest.
Individual roles are often well-matched by only two dimensions, bringing the LLM response distributions within the sampling error of the humans'.
This makes sense because behavior within each game is unique.
The Beauty Contest remains the clearest exception, although the five-dimensional type reduces its distance by 84\% relative to the default.
And using Strategic Sophistication in isolation can improve the fit.
This indicates that dimensions interact, so adding dimensions is not always beneficial.

\subsection{Matching an Individual Human's Behaviors Across Settings}

We now use our method to understand the structure of individual human behavior.
In Section~\ref{sec:match_dist}, we fitted separate mixture weights for each game role to match the distribution of human choices within that role.
Here, we assign each individual a single type to match that individual's choices in every game role that the individual played.
Essentially, we are now fitting the joint distribution over plays in all game roles rather than the marginal distribution for each game role separately.
For this analysis, we use \IndivDecisions decisions made by \IndivSubjects individuals who played at least five game roles.
We first examine how matching quality changes with the number of dimensions, and then study how the fitted types vary across individuals and predict held-out choices.

\subsubsection{Dimensions Needed to Match Individuals}
\label{sec:match-ind}

For each individual, we find the single type that best matches the individual's choices across all game roles they played.
We do this separately for one- through five-dimensional type vectors.
For each candidate type and game role, we calculate the mean absolute difference between the individual's observed choice and ten LLM choices generated by the type's prompt, divided by the game's action range.
We measure the type's matching error for that individual as the average of these game-level distances across the roles the individual played.
For each number of dimensions, we select the type with the lowest matching error.
For example, a matching error of $\varepsilon=0.10$ means that the choices generated by the matched type differ from the individual's observed choices by 10\% of each game's action range on average.

\begin{figure}[h]
\centering
\begin{subfigure}[T]{.4\textwidth}
\includegraphics[width=\linewidth]{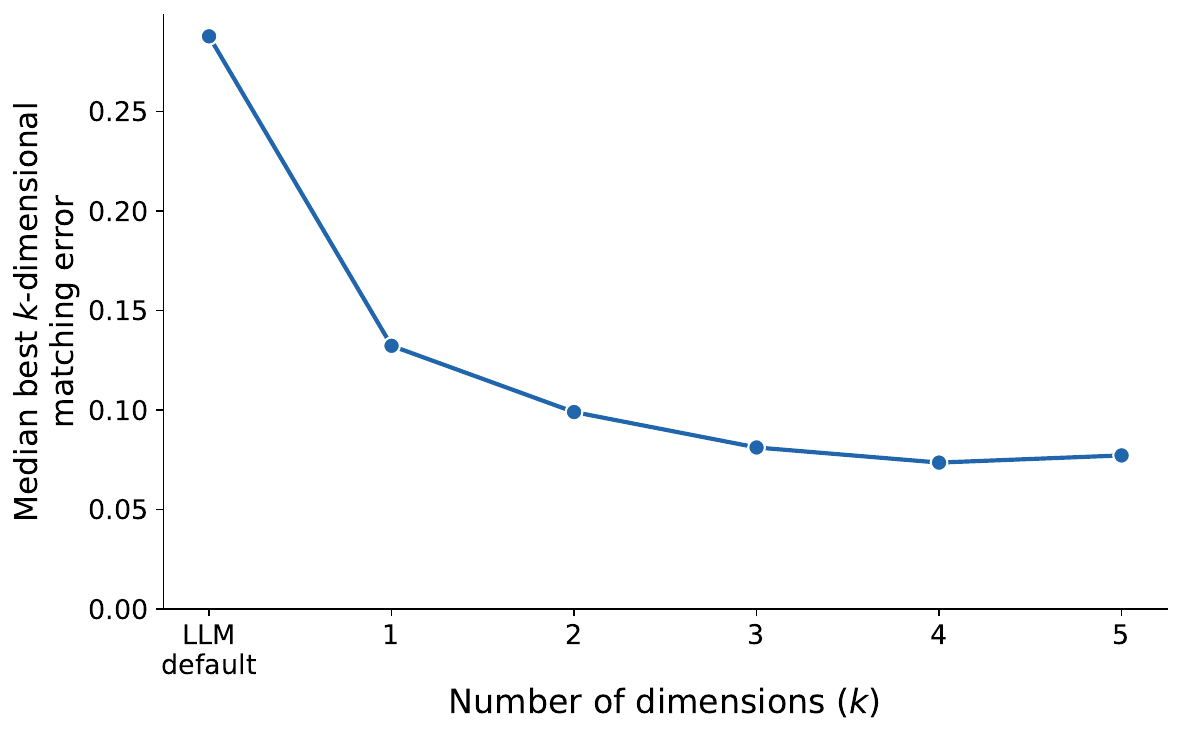}
\end{subfigure}
\begin{subfigure}[T]{.4\textwidth}
\includegraphics[width=\linewidth]{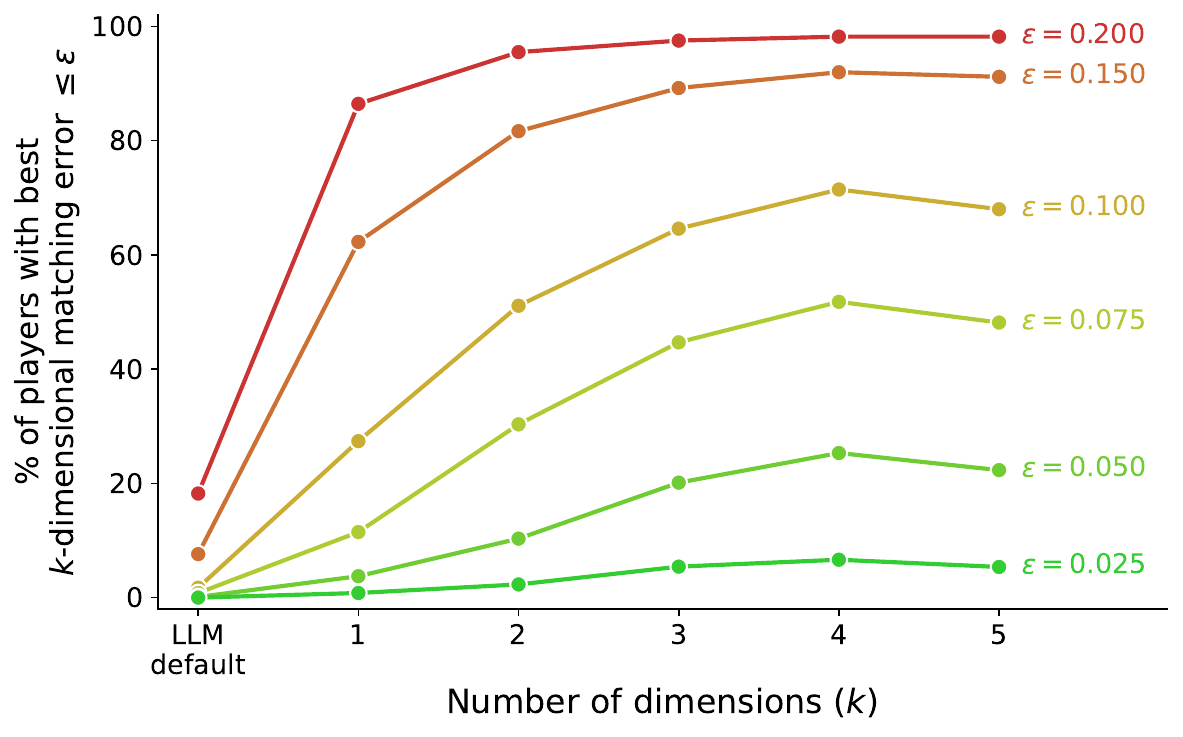}
\end{subfigure}
\caption{Number of dimensions needed to match $N=\IndivSubjects$ individual human players who played at least five game roles.
The left plot shows the best matching distance versus type dimensionality. 
The right plot shows the proportion of players who can be matched by a $k$-dimensional prompt with a matching error no larger than $\varepsilon$.
Each curve corresponds to a matching error threshold $\varepsilon$. }
\label{fig:dimensions}
\end{figure}

Figure~\ref{fig:dimensions} summarizes the individual fits by the number of dimensions. 
The left panel shows that most of the improvement comes from the first three dimensions.
Moving from the default prompt to a one-dimensional type reduces the median matching error from $0.288$ to $0.132$.
As with matching entire distributions in Section~\ref{sec:match_dist}, the fit flattens after three dimensions.
Notably, when matching subjects across games, four dimensions are actually better than five.

The right panel of Figure~\ref{fig:dimensions} reports, for error thresholds $\varepsilon \in \{.025, .05, .075, .1, .15, .2\}$, the proportion of subjects who can be matched by a $k$-dimensional type with error no larger than $\varepsilon$.
The key takeaway is that the population is heterogeneous in how well individuals fit and how many dimensions it takes. 

\subsubsection{Human Heterogeneity and the Distribution of Types}

\begin{figure}[h]
\centering
\includegraphics[width=\linewidth]{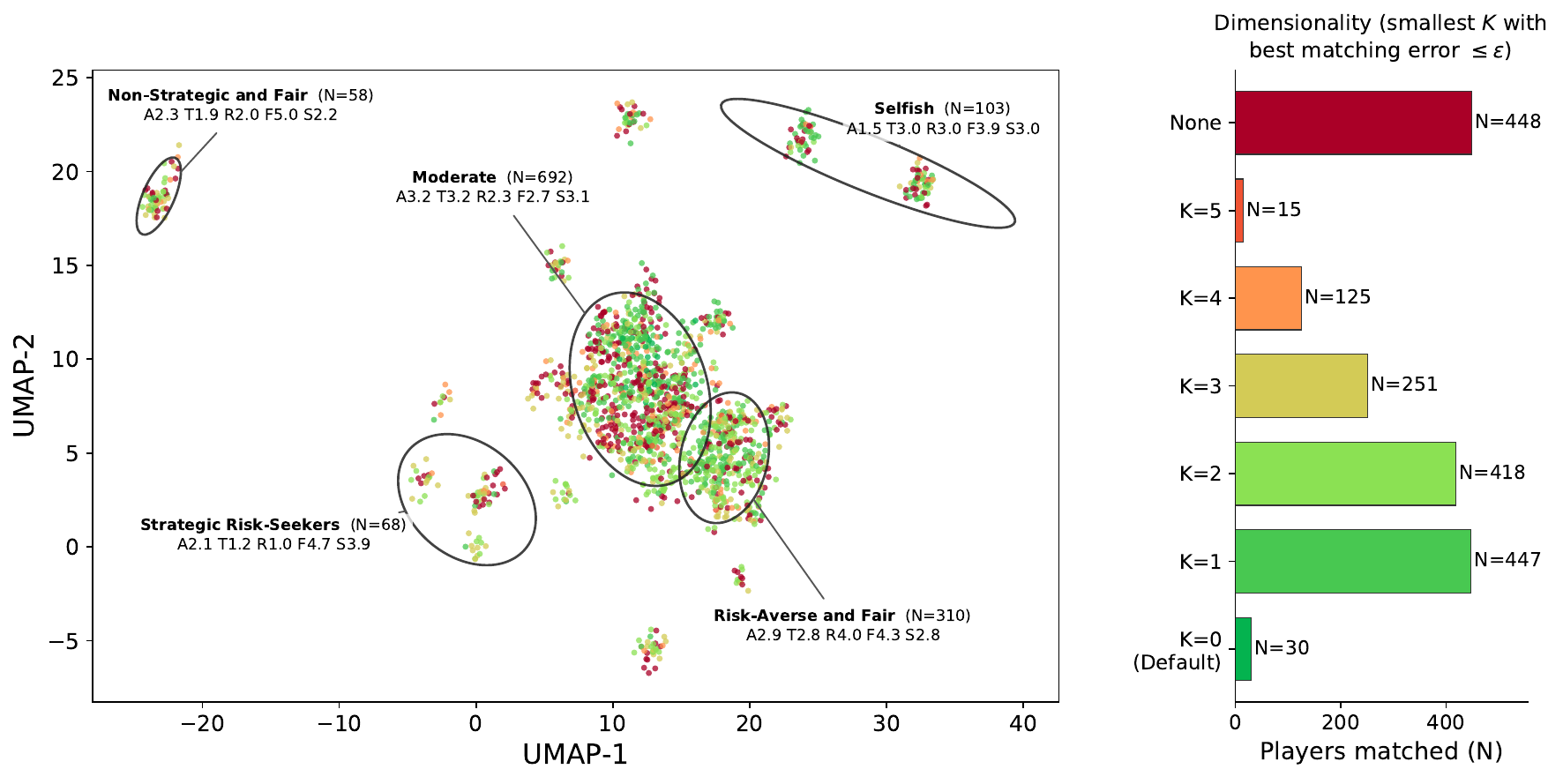}
\caption{Two-dimensional UMAP projections of the best-fitting individual human type vectors. Each dot is a human player represented by a 5-dimensional type vector on a 5-point Likert scale.
Vectors are projected by UMAP. 
Color $k$ indicates the minimal prompt dimensionality to match this player within the error threshold $\varepsilon=0.1$. 
``None'' means no prompt can match this player within the error threshold. 
Only players who have played five or more games are shown ($N=\IndivSubjects$). 
For visual clarity, we add Gaussian random noise ($\sigma=0.5$) to UMAP coordinates.}
\label{fig:heterogeneity}
\end{figure}

Having fit type vectors to each human subject, we can analyze whether there are patterns in the type space.
That is, whether humans cluster into a few tight groups that have fairly similar types, or spread more uniformly across the space. 
To do so, we take each participant's best-fitting five-dimensional type vector and use UMAP, a nonlinear dimension-reduction method \citep{mcinnes2018umap}, to construct a two-dimensional projection of the type vectors.
The left panel of Figure~\ref{fig:heterogeneity} shows this projection, while the right panel reports the smallest number of dimensions needed to match each subject with an error no larger than $\varepsilon=0.1$.

Although people could hypothetically occupy the full space, they cluster around a small set of type vectors.
We see distinctly different types across the clusters.
For instance, the Risk-Averse and Fair types have an average vector of $(A=2.9, T=2.8, R=4.0, F=4.3, S=2.8)$, which is average for most dimensions, but high on Risk Aversion and Fairness. 
The largest ``Moderate'' cluster is close to the middle of the 1-5 range on all dimensions.
Clusters on the left side, including ``Strategic Risk-Seekers'' and ``Non-Strategic and Fair'' types, are generally more risk-loving ($R\le2.0$), emphasizing fairness ($F=5.0$ and $F=4.7$) and low trust ($T\le1.9$).

The right panel shows heterogeneity in how many dimensions are needed to fit different subjects.
Of the \IndivSubjects subjects, 30 are matched by the default prompt within $\varepsilon=0.1$, 447 require one dimension, and so on.
The remaining 448 cannot be matched at this threshold.
Among the subjects who can be matched, nearly 70\% are matched by a type with no more than two dimensions.
So while subjects cluster into fewer than a dozen distinct groups, they are heterogeneous in how many dimensions are needed to fit them well.

\subsubsection{Out-of-Distribution Prediction}

Matching a subject's behaviors across game roles does not establish that the inferred best-fitting type predicts the subject's behavior in new settings.
We therefore examine whether a type inferred only from some subset of games predicts that subject's behavior in another game on which the type is not estimated.\footnote{Here, ``out of distribution'' refers to prediction from our type assignments across game roles that were not used to derive the type.
We do not claim the held-out games are absent from the LLM's training corpus or that performance will generalize.
See \citet{ludwig2026llm} for the assumptions needed to attach broader econometric guarantees to LLM-based predictions.}

We do the following leave-one-game-out exercise for each possible subset of the five dimensions (including all five).
We take each game in turn as the held-out game.
For every subject who played that game and at least four others, we find the type that best matches their choices in those other games.
This becomes the subject's estimated type for that set of dimensions.
We then use the median of the ten choices generated by that type in the held-out game to predict the subject's choice in the held-out game, and measure the absolute difference between the predicted and actual choices, normalized by the range of possible choices in the game.
If several types fit equally well, we average their held-out prediction errors.

This held-out-game exercise would be difficult to implement with traditional dimensionality-reduction methods.
Principal components and factor analysis estimate latent structure among the outcomes included in the estimation data, but predicting behavior in a new setting requires either data from that setting or a dimension-specific model specifying how the latent factors map into choices.
Because the LLM operates in natural language, the inferred type can be applied to a held-out game even when its environment and action space differ from those of the games used to estimate it.

To put our out-of-distribution predictions to a demanding test, we compare them with a benchmark designed to extract as much predictive power as possible from the observed human choices.
For each held-out game, we estimate a histogram-based gradient-boosted tree (HistGBT)~\citep{ke2017lightgbm} that uses each subject's choices in the other games to predict their choice in the held-out game.
We evaluate the model using five-fold cross-validation across subjects, so a subject's own held-out choice is never used to generate that subject's prediction.

These ML predictions are out of sample with respect to the subject but in distribution with respect to the particular game.
The ML benchmark uses the choices of other subjects in the held-out game to estimate how behavior in the other games predicts behavior in that game.
In contrast, our type-based prediction estimates the subject's type from the other games and then applies that type out of distribution to the held-out game.
Thus, the ML benchmark (HistGBT) has the informational advantage of being explicitly trained directly on human behavior in the game it predicts.

\begin{figure}[t]
\centering
\includegraphics[width=\linewidth]{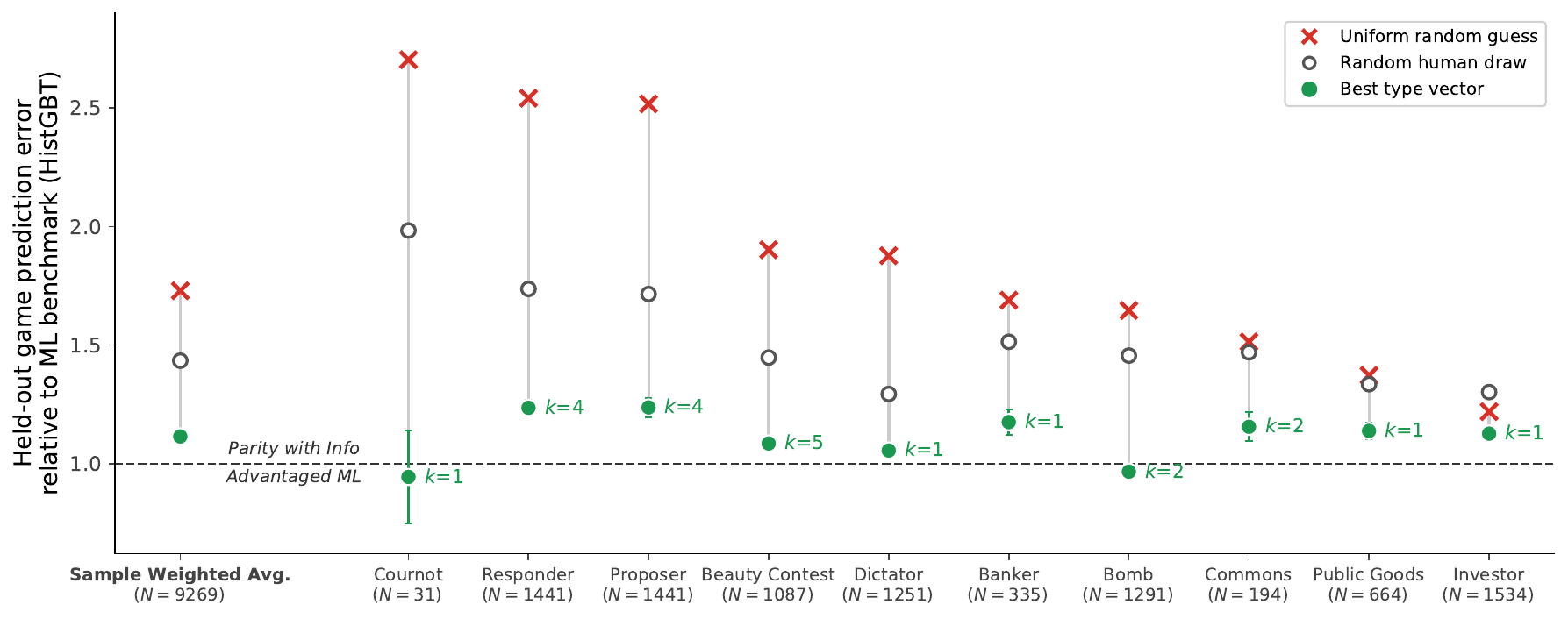}
\caption{Held-out-game prediction error relative to an ML model (HistGBT).
For each game, each point reports the mean absolute error (which is measured as a share of the game's action range), divided by the corresponding HistGBT error.
The dashed line at one marks parity with HistGBT, and lower values indicate lower prediction error.
The green point for each game reports the best-performing type-based prediction, with its label indicating the corresponding dimensionality $k$
The sample-weighted average is the subject-game-weighted mean errors divided by the ML counterparts.
The $N$ beneath each game is its number of subjects and decisions; beneath the average, it is the total number of subject-game observations.
The red x's report the expected error from a uniform random guess, while the open black circle reports the expected error from a randomly selected human choice.
Error bars report $\pm1$ s.e.\ for the type-based mean error divided by the HistGBT mean.}
\label{fig:fit-ratio}
\end{figure}

Figure~\ref{fig:fit-ratio} compares the lowest type-based prediction error, a uniform random guess, and a randomly drawn human choice, all with HistGBT for each game.
Appendix Figure~\ref{fig:fit-hist-ratio} reports results for the selected combination at each dimensionality and the absolute ML error for each game.

The best-performing types produce substantially lower errors than either reference prediction, and approach the information-advantaged ML benchmark.
Taking the best dimensionality for each game, the sample-weighted error is 1.11 times the HistGBT error, compared with 1.43 for a random human draw and 1.73 for a uniform random guess.
Across games, the lowest type-based error ranges from .94 to 1.24 times the HistGBT error.
The point estimates are slightly lower than HistGBT for Bomb and Cournot, although the Cournot estimate is imprecise because only 31 subjects played that game.
In seven of the ten games, one or two dimensions provide the best fit.

\subsection{Alternative Types}
\label{sec:alt-dims}

Although our explorations and analyses so far have focused on prominent economic traits, our approach is not limited to economics. 
Our technique also lets us compare how different traits affect the fit. 
We test two additional sets of traits in matching individual play across the ten game roles.
The first set is five prominent traits from psychology-based theories: the ``OCEAN'' Big 5 personality traits. 
These are Openness, Conscientiousness, Extraversion, Agreeableness, and Neuroticism.
The traits are designed to capture different aspects of human features that can influence behavior, and so represent a natural comparison \citep{costa1992revised,becker2012relationship, jagelka2024economists}

We also explore the performance of atheoretical or ``placebo'' traits.
They are atheoretical in the sense that no plausible relationship between these traits and the way that we should expect humans to play in any of the game roles.
We make up 5 dimensions that seem completely unrelated to any of the game roles. 
These are: Preference for the color orange, Amount of milk in coffee, Interest in cloud shapes, Enjoyment of the sound of rain, and Fondness for the smell of old books.

Our analysis proceeds by repeating the individual matching exercise in Section~\ref{sec:match-ind} separately for the economic, Big 5, and placebo traits.
Each trait takes three Likert values, from 1 to 3.\footnote{We use three rather than five Likert values because the full combination was prohibitively expensive to repeat on additional sets.
Appendix~\ref{app:figs} shows five-point results for five economic dimensions, up to three for the Big 5, and one placebo dimension.}
For every number of dimensions $k$ from one through five, we evaluate every combination of $k$ traits in each set.
Each combination generates $3^k$ prompt types, and we prompt the LLM to respond as each type ten times in every game.
For each trait set and $k$, we select the combination with the lowest median matching error across subjects who played at least five games.

\begin{figure}[h]
\centering
\includegraphics[width=.8\linewidth]{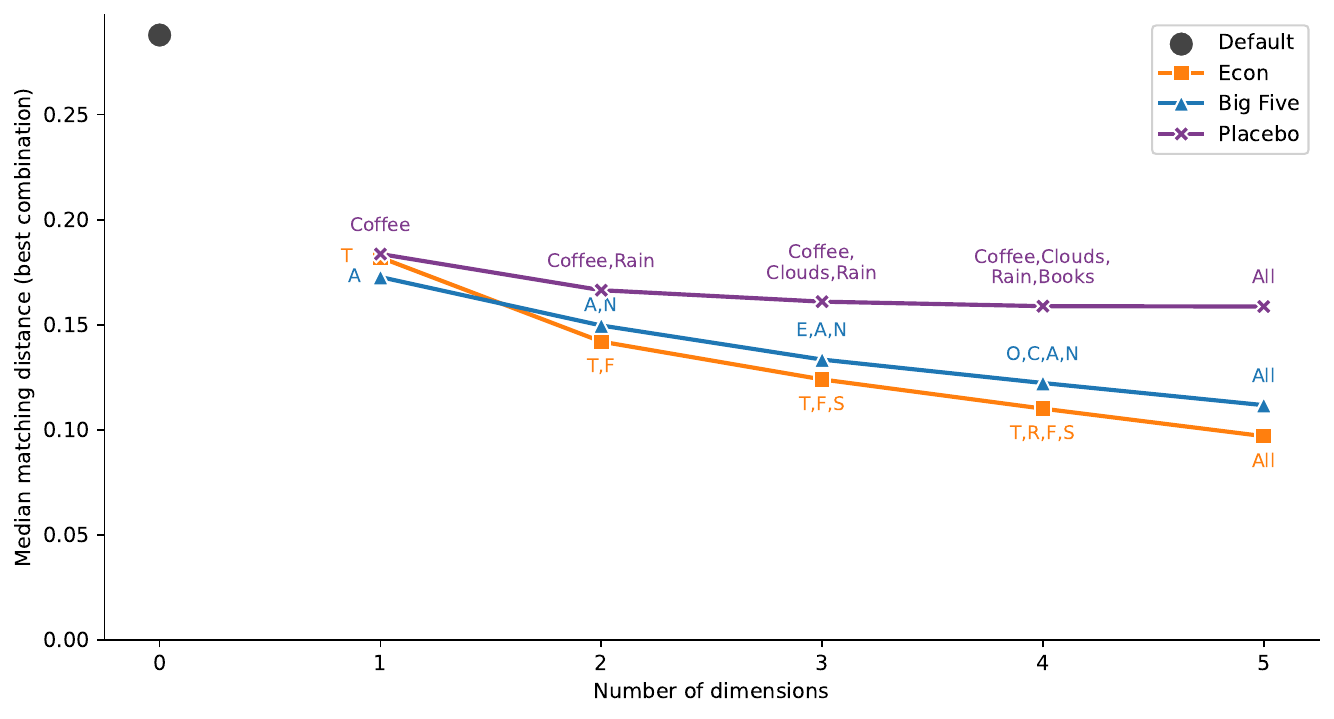}
\caption{Median matching distance across subjects by number of dimensions, for the five economic traits, the Big-5 OCEAN personality traits, and the placebo traits, all on a 3-level Likert scale. 
For each number of dimensions $k$, a point reports the combination of $k$ characteristics from that set with the lowest median matching distance across subjects, where each subject is matched by their best-fitting type vector within that combination. 
The label beside each point names the winning combination by the initial letters of its characteristics.}
\label{fig:alt-key-L3}
\end{figure}

Figure~\ref{fig:alt-key-L3} compares how well each of three trait sets---the five economic traits, the Big 5 personality traits, and the placebos---matches individual subjects.
The horizontal axis reports the number of dimensions, while the vertical axis reports the median matching error across subjects.
For each set and number of dimensions, the figure shows the type that produces the lowest median matching error.

Moving from the default prompt to the best one-dimensional trait substantially improves the fit for all three sets of traits.
Even a placebo trait can help because varying its level induces some behavioral variation, giving the matching procedure possible choices with which to match subjects.
The benefit of adding placebo dimensions, however, quickly diminishes.

The best one-dimensional fit comes from the Big 5's Agreeableness.
One possible interpretation is that Agreeableness is a broad characteristic that combines elements of several of the economic characteristics, including Altruism, Fairness, and Trust.
From two through five dimensions, however, the economic traits provide the best fit.
Indeed, four economic dimensions fit slightly better than all five Big 5 characteristics.
The particular words therefore matter, and they matter in intuitive ways.

This comparison does not establish that our five economic traits are the best possible set.
They achieve most of their reduction in matching error with three dimensions, after which the curve becomes flatter.
For example, another set of traits might produce a more sharply declining curve and achieve the same fit with only two dimensions, or might not flatten out.
Given how much matching error has already been eliminated, however, even the best possible curve is limited in how much further it can reduce the error.
Our ML benchmark suggests that the potential further reduction in error is likely not more than another quarter of the remaining error.

\section{Discussion} \label{sec:discussion}

We began with two related questions: how many dimensions are needed to describe human behavior across settings, and how are people distributed in the resulting space?
Our findings suggest that just a few dimensions generate the substantial heterogeneity we observe within and across games, and the types we identify cluster tightly into distinct groups.

Our broader contribution is to introduce a structured and easy-to-implement modeling method that uses AI together with type vectors to make questions like these empirically tractable across settings and theories.
The types generated by our method are not merely statistical summaries of observed choices.
Once a subject's type is estimated, the same prompt can generate predictions for that subject in any setting described in natural language.
The method therefore allows researchers to compare candidate dimensions and to examine whether types estimated in one set of settings predict behavior in another.

A separate and equally important question concerns the literal interpretation of the traits themselves.
Because an LLM is a black box, changing the component vector value of a characteristic from, say, 3 to 4, may not be affecting the inner workings of the model in ways that correspond to the human notion of that characteristic.
Furthermore, it could be changing several features internally at once \citep{gui2023challenge}.
If so, a nominally low-dimensional type space could still produce a complex and effectively unstructured set of behaviors.
The smooth and generally monotonic effects of the dimensions across games, together with the concentration of subjects around a limited number of types, alleviate this concern.
The dimensions we identify therefore appear to reflect meaningful structure in the behavior generated by the prompts rather than simply the number of values used to label them.
Moreover, even if what the model reads as ``Altruism'' is something else, the type still reproduces the play---as in factor analysis, the count is identified, but the rotation need not be.

We make no claim that traits we study should be interpreted as a test of a particular theory of human behavior. 
Still, the facts that theoretically motivated traits work, and that their relationship to the games seems to match up with existing theories, and some sets perform better than others, suggest that the dimensions do contain information.
Indeed, an exciting avenue of future research is to explore whether this modeling method can be used to evaluate and develop new theories by proposing dimensions and evaluating which can ``best'' approximate human responses with the fewest components.

Another important question for future investigation is how types transfer beyond the game roles used to estimate them.
Our leave-one-game-out exercise shows that types inferred from some games can generalize out of distribution well in our specific context.
Fortunately, the generality of our method makes such questions broadly testable.
The empirical exercises we perform here can be repeated with new dimensions, new populations, and settings that differ substantially from the economic games studied here.
Doing so can show over which domains types remain stable, where additional dimensions are needed, and when types inferred in one domain cease to predict behavior in another, and how populations and settings differ in terms of which theories apply.

\begin{spacing}{0.88}
\bibliographystyle{jpe}
\bibliography{AIreferences}  

@article{harsanyi1967games,
  title={Games with incomplete information played by “Bayesian” players, I--III Part I. The basic model},
  author={Harsanyi, John C},
  journal={Management science},
  volume={14},
  number={3},
  pages={159--182},
  year={1967},
  publisher={INFORMS}
}

@article{samuelson1954pure,
  title={The Pure Theory of Public Expenditure},
  author={Samuelson, Paul A.},
  journal={The Review of Economics and Statistics},
  volume={36},
  number={4},
  pages={387--389},
  year={1954},
  doi={10.2307/1925895}
}

@article{gabaix2014sparsity,
  author  = {Gabaix, Xavier},
  title   = {A Sparsity-Based Model of Bounded Rationality},
  journal = {The Quarterly Journal of Economics},
  year    = {2014},
  volume  = {129},
  number  = {4},
  pages   = {1661--1710},
  doi     = {10.1093/qje/qju024}
}

@article{woodford2020imprecision,
  author  = {Woodford, Michael},
  title   = {Modeling Imprecision in Perception, Valuation, and Choice},
  journal = {Annual Review of Economics},
  year    = {2020},
  volume  = {12},
  pages   = {579--601},
  doi     = {10.1146/annurev-economics-102819-040518}
}

@article{enke2023cognitive,
  author  = {Enke, Benjamin and Graeber, Thomas},
  title   = {Cognitive Uncertainty},
  journal = {The Quarterly Journal of Economics},
  year    = {2023},
  volume  = {138},
  number  = {4},
  pages   = {2021--2067}
}

@article{bordalo2020memory,
  author  = {Bordalo, Pedro and Gennaioli, Nicola and Shleifer, Andrei},
  title   = {Memory, Attention, and Choice},
  journal = {The Quarterly Journal of Economics},
  year    = {2020},
  volume  = {135},
  number  = {3},
  pages   = {1399--1442}
}

@techreport{enke2024attenuation,
  author      = {Enke, Benjamin and Graeber, Thomas and Oprea, Ryan
                 and Yang, Jeffrey},
  title       = {Behavioral Attenuation},
  institution = {National Bureau of Economic Research},
  type        = {Working Paper},
  number      = {32973},
  year        = {2024},
  note        = {Forthcoming in the Journal of Political Economy},
  url         = {https://www.nber.org/papers/w32973}
}

@article{serapiogarcia2025psychometric,
  title={A Psychometric Framework for Evaluating and Shaping Personality Traits in Large Language Models},
  author={Serapio-Garc{\'i}a, Gregory and Safdari, Mustafa and Crepy, Cl{\'e}ment and Sun, Luning and Fitz, Stephen and Romero, Peter and Abdulhai, Marwa and Faust, Aleksandra and Matari{\'c}, Maja},
  journal={Nature Machine Intelligence},
  volume={7},
  pages={1954--1968},
  year={2025},
  doi={10.1038/s42256-025-01115-6}
}

@article{wang2025evaluating,
  title={Evaluating the Ability of Large Language Models to Emulate Personality},
  author={Wang, Yilei and Zhao, Jiabao and Ones, Deniz S. and He, Liang and Xu, Xin},
  journal={Scientific Reports},
  volume={15},
  pages={519},
  year={2025},
  doi={10.1038/s41598-024-84109-5}
}

@article{huang2026designing,
  title={Designing {AI}-Agents With Personalities: A Psychometric Approach},
  author={Huang, Muhua and Zhang, Xijuan and Soto, Christopher and Evans, James},
  journal={Personality Science},
  volume={7},
  pages={1--26},
  year={2026},
  doi={10.1177/27000710251406471}
}

@article{sakai2026effects,
  title={Effects of Personality Steering on Cooperative Behavior in Large Language Model Agents},
  author={Sakai, Mizuki and Yokoyama, Mizuki and Tateishi, Wakaba and Ichinose, Genki},
  journal={Scientific Reports},
  volume={16},
  pages={25076},
  year={2026},
  doi={10.1038/s41598-026-56163-8}
}

@book{keynes1936general,
  title={The General Theory of Employment, Interest and Money},
  author={Keynes, John Maynard},
  year={1936},
  publisher={Macmillan},
  address={London}
}

@book{lloyd1833two,
  title={Two Lectures on the Checks to Population: Delivered Before the University of Oxford, in Michaelmas Term 1832},
  author={Lloyd, William Forster},
  year={1833},
  publisher={S. Collingwood},
  address={Oxford}
}

@article{mcinnes2018umap,
    author = {McInnes, Leland and Healy, John and Saul, Nathaniel and Gro{\ss}berger, Lukas},
    title = {UMAP: Uniform Manifold Approximation and Projection},
    journal = {Journal of Open Source Software},
    year = {2018},
    volume = {3},
    number = {29},
    pages = {861},
    doi = {10.21105/joss.00861}
  }

@article{forsythe1994fairness,
  title={Fairness in Simple Bargaining Experiments},
  author={Forsythe, Robert and Horowitz, Joel L. and Savin, N. E. and Sefton, Martin},
  journal={Games and Economic Behavior},
  volume={6},
  number={3},
  pages={347--369},
  year={1994},
  doi={10.1006/game.1994.1021}
}

@article{andreoni1995cooperation,
  title={Cooperation in Public-Goods Experiments: Kindness or Confusion?},
  author={Andreoni, James},
  journal={American Economic Review},
  volume={85},
  number={4},
  pages={891--904},
  year={1995}
}

@article{crosetto2013bomb,
  title={The ``Bomb'' Risk Elicitation Task},
  author={Crosetto, Paolo and Filippin, Antonio},
  journal={Journal of Risk and Uncertainty},
  volume={47},
  number={1},
  pages={31--65},
  year={2013},
  doi={10.1007/s11166-013-9170-z}
}

@article{walker1990rent,
  title={Rent Dissipation in a Limited-Access Common-Pool Resource: Experimental Evidence},
  author={Walker, James M. and Gardner, Roy and Ostrom, Elinor},
  journal={Journal of Environmental Economics and Management},
  volume={19},
  number={3},
  pages={203--211},
  year={1990},
  doi={10.1016/0095-0696(90)90069-B}
}

@article{huck1999learning,
  title={Learning in Cournot Oligopoly---An Experiment},
  author={Huck, Steffen and Normann, Hans-Theo and Oechssler, J{\"o}rg},
  journal={The Economic Journal},
  volume={109},
  number={454},
  pages={C80--C95},
  year={1999},
  doi={10.1111/1468-0297.00418}
}

@article{fudenberg2022measuring,
  title={Measuring the completeness of economic models},
  author={Fudenberg, Drew and Kleinberg, Jon and Liang, Annie and Mullainathan, Sendhil},
  journal={Journal of Political Economy},
  volume={130},
  number={4},
  pages={956--990},
  year={2022},
  publisher={The University of Chicago Press Chicago, IL}
}

@article{krakauer2024complex,
  title={The complex world},
  author={Krakauer, David C},
  journal={Santa Fe Institute Press, Santa Fe, NM},
  year={2024}
}

@article{akata2025playing,
  title={Playing repeated games with large language models},
  author={Akata, Elif and Schulz, Lion and Coda-Forno, Julian and Oh, Seong Joon and Bethge, Matthias and Schulz, Eric},
  journal={Nature Human Behaviour},
  volume={9},
  number={7},
  pages={1380--1390},
  year={2025},
  publisher={Nature Publishing Group UK London}
}

@article{ashokkumar2026large,
  title={Large language models can predict the results of social science experiments},
  author={Ashokkumar, Ashwini and Hewitt, Luke and Ghezae, Isaias and Willer, Robb},
  journal={Nature},
  pages={1--8},
  year={2026},
  publisher={Nature Publishing Group UK London}
}

@article{jackson2025ai,
  title={Ai behavioral science},
  author={Jackson, Matthew O and Me, Qiaozhu and Wang, Stephanie W and Xie, Yutong and Yuan, Walter and Benzell, Seth and Brynjolfsson, Erik and Camerer, Colin F and Evans, James and Jabarian, Brian and others},
  journal={arXiv preprint arXiv:2509.13323},
  year={2025}
}

@article{ludwig2024machine,
  title={Machine learning as a tool for hypothesis generation},
  author={Ludwig, Jens and Mullainathan, Sendhil},
  journal={The Quarterly Journal of Economics},
  volume={139},
  number={2},
  pages={751--827},
  year={2024},
  publisher={Oxford University Press}
}

@article{anthis2025position,
  title={Position: Llm social simulations are a promising research method},
  author={Anthis, Jacy Reese and Liu, Ryan and Richardson, Sean M and Kozlowski, Austin C and Koch, Bernard and Brynjolfsson, Erik and Evans, James and Bernstein, Michael S},
  journal={Forty-second International Conference on Machine Learning Position Paper Track},
  year={2025}
}

@misc{hullman2026involve,
      title={This human study did not involve human subjects: Validating LLM simulations as behavioral evidence}, 
      author={Jessica Hullman and David Broska and Huaman Sun and Aaron Shaw},
      year={2026},
      eprint={2602.15785},
      archivePrefix={arXiv},
      primaryClass={cs.AI},
      url={https://arxiv.org/abs/2602.15785}, 
}

@article{binz2023using,
  title={Using cognitive psychology to understand GPT-3},
  author={Binz, Marcel and Schulz, Eric},
  journal={Proceedings of the National Academy of Sciences},
  volume={120},
  number={6},
  pages={e2218523120},
  year={2023},
  publisher={National Academy of Sciences}
}

@article{charness2002understanding,
  title     = {Understanding social preferences with simple tests},
  author    = {Charness, Gary and Rabin, Matthew},
  journal   = {The quarterly journal of economics},
  volume    = {117},
  number    = {3},
  pages     = {817--869},
  year      = {2002},
  publisher = {Oxford University Press}
}

@article{chapman2023econographics,
  author  = {Chapman, Jonathan and Dean, Mark and Ortoleva, Pietro and Snowberg, Erik and Camerer, Colin F.},
  title   = {Econographics},
  journal = {Journal of Political Economy Microeconomics},
  year    = {2023},
  volume  = {1},
  number  = {1},
  pages   = {115--161},
  doi     = {10.1086/723044}
}

@article{stango2023taxonomy,
  author  = {Stango, Victor and Zinman, Jonathan},
  title   = {We Are All Behavioural, More or Less: A Taxonomy of Consumer Decision-Making},
  journal = {The Review of Economic Studies},
  year    = {2023},
  volume  = {90},
  number  = {3},
  pages   = {1470--1498},
  doi     = {10.1093/restud/rdac055}
}

@article{dean2019empirical,
  author  = {Dean, Mark and Ortoleva, Pietro},
  title   = {The Empirical Relationship between Nonstandard Economic Behaviors},
  journal = {Proceedings of the National Academy of Sciences},
  year    = {2019},
  volume  = {116},
  number  = {33},
  pages   = {16262--16267},
  doi     = {10.1073/pnas.1821353116}
}

@article{zhu2025capturing,
  author  = {Zhu, Jian-Qiao and Peterson, Joshua C. and
             Enke, Benjamin and Griffiths, Thomas L.},
  title   = {Capturing the Complexity of Human Strategic
             Decision-Making with Machine Learning},
  journal = {Nature Human Behaviour},
  year    = {2025},
  volume  = {9},
  number  = {10},
  pages   = {2114--2120},
  doi     = {10.1038/s41562-025-02230-5}
}

@article{binz2025foundation,
  author  = {Binz, Marcel and Akata, Elif and Bethge, Matthias and
             Br{\"a}ndle, Franziska and Callaway, Fred and
             Coda-Forno, Julian and Dayan, Peter and Demircan, Can and
             Eckstein, Maria K. and {\'E}ltet{\H{o}}, No{\'e}mi and
             Griffiths, Thomas L. and Haridi, Susanne and
             Jagadish, Akshay K. and Ji-An, Li and Kipnis, Alexander and
             Kumar, Sreejan and Ludwig, Tobias and Mathony, Marvin and
             Mattar, Marcelo and Modirshanechi, Alireza and
             Nath, Surabhi S. and Peterson, Joshua C. and Rmus, Milena and
             Russek, Evan M. and Saanum, Tankred and
             Schubert, Johannes A. and Schulze Buschoff, Luca M. and
             Singhi, Nishad and Sui, Xin and Thalmann, Mirko and
             Theis, Fabian J. and Truong, Vuong and Udandarao, Vishaal and
             Voudouris, Konstantinos and Wilson, Robert and
             Witte, Kristin and Wu, Shuchen and Wulff, Dirk U. and
             Xiong, Huadong and Schulz, Eric},
  title   = {A Foundation Model to Predict and Capture Human Cognition},
  journal = {Nature},
  year    = {2025},
  volume  = {644},
  pages   = {1002--1009},
  doi     = {10.1038/s41586-025-09215-4}
}

@article{liang2026using,
  author  = {Liang, Annie},
  title   = {Using Machine Learning to Generate, Clarify,
             and Improve Economic Models},
  journal = {Journal of Economic Literature},
  year    = {2026},
  note    = {Accepted},
  url     = {https://anniehliang.com/papers/JEL.pdf}
}

@article{fudenberg2019predicting,
  author  = {Fudenberg, Drew and Liang, Annie},
  title   = {Predicting and Understanding Initial Play},
  journal = {American Economic Review},
  year    = {2019},
  volume  = {109},
  number  = {12},
  pages   = {4112--4141},
  doi     = {10.1257/aer.20180654}
}

@article{fudenberg2006advancing,
  author  = {Fudenberg, Drew},
  title   = {Advancing Beyond Advances in Behavioral Economics},
  journal = {Journal of Economic Literature},
  year    = {2006},
  volume  = {44},
  number  = {3},
  pages   = {694--711},
  doi     = {10.1257/jel.44.3.694}
}

@article{becker2012relationship,
  author  = {Becker, Anke and Deckers, Thomas and Dohmen, Thomas and Falk, Armin and Kosse, Fabian},
  title   = {The Relationship between Economic Preferences and Psychological Personality Measures},
  journal = {Annual Review of Economics},
  year    = {2012},
  volume  = {4},
  pages   = {453--478},
  doi     = {10.1146/annurev-economics-080511-110922}
}

@article{jagelka2024economists,
  author  = {Jagelka, Tom{\'a}{\v{s}}},
  title   = {Are Economists' Preferences Psychologists' Personality Traits? A Structural Approach},
  journal = {Journal of Political Economy},
  year    = {2024},
  volume  = {132},
  number  = {3},
  pages   = {910--970},
  doi     = {10.1086/726908}
}

@article{bruhin2019many,
  author  = {Bruhin, Adrian and Fehr, Ernst and Schunk, Daniel},
  title   = {The Many Faces of Human Sociality: Uncovering the Distribution and Stability of Social Preferences},
  journal = {Journal of the European Economic Association},
  year    = {2019},
  volume  = {17},
  number  = {4},
  pages   = {1025--1069},
  doi     = {10.1093/jeea/jvy018}
}

@article{peterson2021using,
  title={Using large-scale experiments and machine learning to discover theories of human decision-making},
  author={Peterson, Joshua C and Bourgin, David D and Agrawal, Mayank and Reichman, Daniel and Griffiths, Thomas L},
  journal={Science},
  volume={372},
  number={6547},
  pages={1209--1214},
  year={2021},
  publisher={American Association for the Advancement of Science}
}

@misc{qian2026strategicAI,
      author = {Qian, Crystal and Zhu, Kehang and Horton, John and Manning, Benjamin and Tsai, Vivian and Wexler, James and Thain, Nithum},
      title = {Strategic Tradeoffs Between Humans and AI in Multi-Agent Bargaining},
      year = {2026},
      isbn = {9798400719844},
      publisher = {Association for Computing Machinery},
      address = {New York, NY, USA},
      url = {https://doi.org/10.1145/3742413.3789078},
      doi = {10.1145/3742413.3789078},
      booktitle = {Proceedings of the 31st International Conference on Intelligent User Interfaces},
      pages = {1625–1646},
      numpages = {22},
      series = {IUI '26}
}

@misc{peng2026funhouse,
      title={Digital Twins as Funhouse Mirrors: Five Key Distortions}, 
      author={Tianyi Peng and George Gui and Melanie Brucks and Daniel J. Merlau and Grace Jiarui Fan and Malek Ben Sliman and Eric J. Johnson and Abdullah Althenayyan and Silvia Bellezza and Dante Donati and Hortense Fong and Elizabeth Friedman and Ariana Guevara and Mohamed Hussein and Kinshuk Jerath and Bruce Kogut and Akshit Kumar and Kristen Lane and Hannah Li and Vicki Morwitz and Oded Netzer and Patryk Perkowski and Olivier Toubia},
      year={2026},
      eprint={2509.19088},
      archivePrefix={arXiv},
      primaryClass={cs.CY},
      url={https://arxiv.org/abs/2509.19088}, 
}

@techreport{horton2023large,
  title        = {Large Language Models as Simulated Economic Agents: What Can We Learn from Homo Silicus?},
  author       = {Horton, John J and Filippas, Apostolos and Manning, Benjamin S},
  institution  = {National Bureau of Economic Research},
  type         = {Working Paper},
  series       = {Working Paper Series},
  number       = {31122},
  year         = {2023},
  month        = {February},
  doi          = {10.3386/w31122},
  URL          = {http://www.nber.org/papers/w31122},
}

@unpublished{ellis2026predicting,
  author = {Ellis, Keaton and Kariv, Shachar and Ozbay, Erkut Y.},
  title  = {Predicting and Understanding Individual-Level
            Choice Under Risk},
  year   = {2026},
  note   = {Working paper},
  url    = {https://eml.berkeley.edu/~kariv/EKO_I.pdf}
}

@article{ludwig2026llm,
  title={Large Language Models: An Applied Econometric Framework},
  author={Ludwig, Jens and Mullainathan, Sendhil and Rambachan, Ashesh},
  journal={Annual Review of Economics},
  volume={18},
  year={2026},
  note={Forthcoming. NBER Working Paper 33344}
}

@misc{mullainathan2025anomalies,
      title={From Predictive Algorithms to Automatic Generation of Anomalies}, 
      author={Sendhil Mullainathan and Ashesh Rambachan},
      year={2025},
      eprint={2404.10111},
      archivePrefix={arXiv},
      primaryClass={econ.EM},
      url={https://arxiv.org/abs/2404.10111}, 
}

@article{vafa2024world,
    title={Evaluating the World Model Implicit in a Generative Model},
    author={Vafa, Keyon and Chen, Justin Y and Rambachan, Ashesh and Kleinberg, Jon and Mullainathan, Sendhil},
    journal={Neural Information Processing Systems},
    year={2024},
}

@article{guth1982experimental,
  title={An experimental analysis of ultimatum bargaining},
  author={G{\"u}th, Werner and Schmittberger, Rolf and Schwarze, Bernd},
  journal={Journal of Economic Behavior \& Organization},
  volume={3}, number={4}, pages={367--388}, year={1982},
  publisher={Elsevier}
}

@article{berg1995trust,
  title={Trust, reciprocity, and social history},
  author={Berg, Joyce and Dickhaut, John and McCabe, Kevin},
  journal={Games and Economic Behavior},
  volume={10}, number={1}, pages={122--142}, year={1995},
  publisher={Elsevier}
}

@article{andreoni2002giving,
  title={Giving according to {GARP}: An experimental test of the consistency of preferences for altruism},
  author={Andreoni, James and Miller, John},
  journal={Econometrica},
  volume={70}, number={2}, pages={737--753}, year={2002},
  publisher={Wiley}
}

@article{gui2023challenge,
  title={The challenge of using {LLMs} to simulate human behavior: A causal inference perspective},
  author={Gui, George and Toubia, Olivier},
  journal={arXiv preprint arXiv:2312.15524},
  year={2023}
}

@article{nagel1995unraveling,
  title={Unraveling in guessing games: An experimental study},
  author={Nagel, Rosemarie},
  journal={The American Economic Review},
  volume={85}, number={5}, pages={1313--1326}, year={1995}
}

@article{camerer2004cognitive,
  title={A cognitive hierarchy model of games},
  author={Camerer, Colin F and Ho, Teck-Hua and Chong, Juin-Kuan},
  journal={The Quarterly Journal of Economics},
  volume={119}, number={3}, pages={861--898}, year={2004},
  publisher={Oxford University Press}
}

@inproceedings{aher2023using,
  title={Using large language models to simulate multiple humans and replicate human subject studies},
  author={Aher, Gati V and Arriaga, Rosa I and Kalai, Adam Tauman},
  booktitle={International Conference on Machine Learning},
  pages={337--371},
  year={2023},
  organization={PMLR}
}

@techreport{abdel2026life,
 title = "Revealing Life Preferences Through LLMs",
 author = "Abdel Haq, Omar and Chandra, Amitabh and Jagelka, Tomáš and Luttmer, Erzo F.P and Schwartzstein, Joshua",
 institution = "National Bureau of Economic Research",
 type = "Working Paper",
 series = "Working Paper Series",
 number = "35185",
 year = "2026",
 month = "May",
 doi = {10.3386/w35185},
 URL = "http://www.nber.org/papers/w35185",
}

@unpublished{gao2026llm,
  author = {Wayne Gao and Sukjin Han and Annie Liang},
  title = {How Well Do {LLMs} Predict Human Behavior? {A} Measure of their Pretrained Knowledge},
  year = {2026},
  note = {Working paper}
}

@book{camerer2003behavioral,
  title={Behavioral game theory: Experiments in strategic interaction},
  author={Camerer, Colin F},
  year={2003},
  publisher={Princeton university press}
}

@article{kahneman1979prospect,
  title={Prospect theory: An analysis of decisions under risk},
  author={Kahneman, Daniel and Tversky, Amos},
  journal={Econometrica},
  volume={47},
  pages={278},
  year={1979}
}

@article{laibson1997golden,
  title={Golden eggs and hyperbolic discounting},
  author={Laibson, David},
  journal={The Quarterly Journal of Economics},
  volume={112},
  number={2},
  pages={443--478},
  year={1997},
  publisher={MIT Press}
}

@article{fehr1999theory,
  title={A theory of fairness, competition, and cooperation},
  author={Fehr, Ernst and Schmidt, Klaus M},
  journal={The quarterly journal of economics},
  volume={114},
  number={3},
  pages={817--868},
  year={1999},
  publisher={MIT press}
}

@article{bolton2000erc,
  title={ERC: A theory of equity, reciprocity, and competition},
  author={Bolton, Gary E and Ockenfels, Axel},
  journal={American economic review},
  volume={91},
  number={1},
  pages={166--193},
  year={2000},
  publisher={American Economic Association}
}

@article{xie2025using,
  title={Using large language models to categorize strategic situations and decipher motivations behind human behaviors},
  author={Xie, Yutong and Mei, Qiaozhu and Yuan, Walter and Jackson, Matthew O},
  journal={Proceedings of the National Academy of Sciences},
  volume={122},
  number={35},
  pages={e2512075122},
  year={2025},
  publisher={National Academy of Sciences}
}

@book{costa1992revised,
  author    = {Costa, Paul T. and McCrae, Robert R.},
  title     = {Revised {NEO} Personality Inventory ({NEO PI-R}) and {NEO} Five-Factor Inventory ({NEO-FFI})},
  year      = {1992},
  publisher = {Psychological Assessment Resources},
  address   = {Odessa, FL}
}

@article{chen2023emergence,
  title={The emergence of economic rationality of GPT},
  author={Chen, Yiting and Liu, Tracy Xiao and Shan, You and Zhong, Songfa},
  journal={Proceedings of the National Academy of Sciences},
  volume={120},
  number={51},
  pages={e2316205120},
  year={2023},
  publisher={National Academy of Sciences}
}

@techreport{manning2024automated,
  title={Automated social science: Language models as scientist and subjects},
  author={Manning, Benjamin S and Zhu, Kehang and Horton, John J},
  year={2024},
  institution={National Bureau of Economic Research}
}

@article{dillion2023can,
  title={Can AI language models replace human participants?},
  author={Dillion, Danica and Tandon, Niket and Gu, Yuling and Gray, Kurt},
  journal={Trends in Cognitive Sciences},
  volume={27},
  number={7},
  pages={597--600},
  year={2023},
  publisher={Elsevier}
}

@article{argyle2023out,
  title={Out of one, many: Using language models to simulate human samples},
  author={Argyle, Lisa P and Busby, Ethan C and Fulda, Nancy and Gubler, Joshua R and Rytting, Christopher and Wingate, David},
  journal={Political Analysis},
  volume={31},
  number={3},
  pages={337--351},
  year={2023},
  publisher={Cambridge University Press}
}

@article{kim2023ai,
  title={Ai-augmented surveys: Leveraging large language models and surveys for opinion prediction},
  author={Kim, Junsol and Lee, Byungkyu},
  journal={arXiv preprint arXiv:2305.09620},
  year={2023}
}

@article{kozlowski2024silico,
  title={In silico sociology: forecasting COVID-19 polarization with large language models},
  author={Kozlowski, Austin C and Kwon, Hyunku and Evans, James A},
  journal={arXiv preprint arXiv:2407.11190},
  year={2024}
}

@misc{kozlowski_evans_2025, title={Simulating Subjects:  The Promise and Peril of AI Stand-ins for Social Agents and Interactions}, url={osf.io/preprints/socarxiv/vp3j2_v3}, DOI={10.31235/osf.io/vp3j2_v3}, publisher={SocArXiv}, author={Kozlowski, Austin C and Evans, James}, year={2025}, month={Apr}}

@article{lippert2024can,
  title={Can large language models help predict results from a complex behavioural science study?},
  author={Lippert, Steffen and Dreber, Anna and Johannesson, Magnus and Tierney, Warren and Cyrus-Lai, Wilson and Uhlmann, Eric Luis and Emotion Expression Collaboration and Pfeiffer, Thomas},
  journal={Royal Society Open Science},
  volume={11},
  number={9},
  pages={240682},
  year={2024},
  publisher={The Royal Society}
}

@book{cournot1838recherches,
  title={Recherches sur les principes math{\'e}matiques de la th{\'e}orie des richesses},
  author={Cournot, Augustin},
  year={1838},
  publisher={L. Hachette},
  address={Paris}
}

@article{abdurahman2024perils,
  title={Perils and opportunities in using large language models in psychological research},
  author={Abdurahman, Suhaib and Atari, Mohammad and Karimi-Malekabadi, Farzan and Xue, Mona J and Trager, Jackson and Park, Peter S and Golazizian, Preni and Omrani, Ali and Dehghani, Morteza},
  journal={PNAS nexus},
  volume={3},
  number={7},
  pages={245},
  year={2024},
  publisher={Oxford University Press US}
}

@article{yeykelis2024using,
  title={Using Large Language Models to Create AI Personas for Replication, Generalization and Prediction of Media Effects: An Empirical Test of 133 Published Experimental Research Findings},
  author={Yeykelis, Leo and Pichai, Kaavya and Cummings, James J and Reeves, Byron},
  journal={arXiv preprint arXiv:2408.16073},
  year={2024}
}

@article{park2023,
  title={Generative Agents: Interactive Simulacra of Human Behavior},
  author={Park, Joon Sung and O'Brien, Joseph and and Cai, Carrie Jun and Morris, Meredith Ringel and Liang, Percy and Bernstein, Michael S.},
  journal={UIST '23: Proceedings of the 36th Annual ACM Symposium on User Interface Software and Technology},
  year={2023}
}

@article{park2024generative,
  title={Generative agent simulations of 1,000 people},
  author={Park, Joon Sung and Zou, Carolyn Q and Shaw, Aaron and Hill, Benjamin Mako and Cai, Carrie and Morris, Meredith Ringel and Willer, Robb and Liang, Percy and Bernstein, Michael S},
  journal={arXiv preprint arXiv:2411.10109},
  year={2024}
}

@misc{manning2025general,
      title={General Social Agents}, 
      author={Benjamin S. Manning and John J. Horton},
      institution = {National Bureau of Economic Research},
      type = {Working Paper},
      series = {Working Paper Series},
      number = {34937},
      year = {2026},
      month = {March},
      doi = {10.3386/w34937},
      url = {http://www.nber.org/papers/w34937},
}

@article{ross2024llm,
  title={Llm economicus? mapping the behavioral biases of llms via utility theory},
  author={Ross, Jillian and Kim, Yoon and Lo, Andrew W},
  journal={arXiv preprint arXiv:2408.02784},
  year={2024}
}

@unpublished{andrews2025transfer,
  author = {Isaiah Andrews and Drew Fudenberg and Lihua Lei and Annie Liang and Chaofeng Wu},
  title = {The Transfer Performance of Economic Models},
  year = {2025},
  note = {Working paper}
}

@article{holt2002risk,
Author = {Holt, Charles A. and Laury, Susan K.},
Title = {Risk Aversion and Incentive Effects },
Journal = {American Economic Review},
Volume = {92},
Number = {5},
Year = {2002},
Month = {December},
Pages = {1644–1655},
DOI = {10.1257/000282802762024700}
}

@article{ke2017lightgbm,
  title={Lightgbm: A highly efficient gradient boosting decision tree},
  author={Ke, Guolin and Meng, Qi and Finley, Thomas and Wang, Taifeng and Chen, Wei and Ma, Weidong and Ye, Qiwei and Liu, Tie-Yan},
  journal={Advances in neural information processing systems},
  volume={30},
  year={2017}
}

@article{lin2020evidence,
  title={Evidence of general economic principles of bargaining and trade from 2,000 classroom experiments},
  author={Lin, Po-Hsuan and Brown, Alexander L. and Imai, Taisuke and Wang, Joseph Tao-yi and Wang, Stephanie W. and Camerer, Colin F.},
  journal={Nature Human Behaviour},
  volume={4},
  pages={917--927},
  year={2020}
}
\end{spacing}

\newpage \clearpage
\appendix

\renewcommand{\thefigure}{A\arabic{figure}} 
\setcounter{figure}{0}  

\renewcommand{\thetable}{A\arabic{table}} 
\setcounter{table}{0}  

\setcounter{page}{1}

\begin{center}
{\bf  Supplemental Appendix for  \\
``\PaperTitle''\\
by Jackson, Manning, Xie, Yuan, and Mei. }  
\end{center}

\onehalfspacing

\section{AI Prompts and Game Instructions}
\label{app:prompts}

Sections~\ref{sec:prompt-method} and~\ref{sec:game-data} describe how we pair each type vector with the instructions for a game role.
Here, we give the exact prompts used to generate the AI choices.
We also describe the LLMs and data pipeline used to generate our data.

\subsection{Type-vector and Default Prompts}

Each elicitation pairs one prompt with the instructions for one game role.
The type-vector or default prompt occupies the system role, and the game instructions occupy the user role.
Our main analysis uses \texttt{gpt-4.1} through the Azure Batch deployment\footnote{Azure OpenAI batch deployments: \url{https://learn.microsoft.com/en-us/azure/foundry/openai/how-to/batch}, retrieved on Aug 17, 2026. } and collects ten valid responses (i.e., responses that can be parsed into choices within the action range specified by game instructions) for every pairing of a prompt and game role.
We use the default API hyperparameters and do not specify temperature, top-$p$, or penalty parameters.

For a type vector with $k$ characteristics measured on a scale with $L$ Likert levels, we construct the prompt by inserting the selected characteristic names and values, along with the number of scale levels $L$, into the following template.

\vspace{3mm}
\begin{promptpart}{personacol}{Type-vector prompt template}
You are a player characterized by the following profile (each dimension is measured on a Likert scale, where 1 is the lowest level and $[$L$]$ is the highest level):
\begin{itemize}[leftmargin=1.2em,nosep,topsep=3pt]
    \item $[$Dimension 1$]$: $[$Value 1$]$ out of $[$L$]$\\
    ...
    \item $[$Dimension K$]$: $[$Value K$]$ out of $[$L$]$
\end{itemize}
\end{promptpart}
\vspace{3mm}

For example, the type vector $(2,4)$ over Altruism and Risk Aversion produces the following prompt when $L=5$.

\vspace{3mm}
\begin{promptpart}{personacol}{Type-vector prompt for $(A,R)=(2,4)$}
You are a player characterized by the following profile (each dimension is measured on a Likert scale, where 1 is the lowest level and 5 is the highest level):
\begin{itemize}[leftmargin=1.2em,nosep,topsep=3pt]
\item Altruism: 2 out of 5
\item Risk Aversion: 4 out of 5
\end{itemize}
\end{promptpart}
\vspace{3mm}

The default system prompt uses the exact text shown below and the same game instructions.

\vspace{3mm}
\begin{promptpart}{personacol}{Default prompt}
You are a helpful assistant.
\end{promptpart}
\vspace{3mm}

\paragraph{Characteristics.}
The five economic characteristics used in our main analysis are Altruism, Trust, Risk Aversion, Fairness, and Strategic Sophistication.
For comparison, we also use the Big 5 personality characteristics: Openness, Conscientiousness, Extraversion, Agreeableness, and Neuroticism.
The placebo characteristics are Preference for the color orange, Amount of milk in coffee, Interest in cloud shapes, Enjoyment of the sound of rain, and Fondness for the smell of old books.

\subsection{Game Instructions}
\label{app:game-instructions}

The boxes below reproduce exactly the game instructions supplied for each game role.
The wording, capitalization, punctuation, and paragraph breaks are unchanged.
The games are based on standard dictator and ultimatum games \citep{guth1982experimental,forsythe1994fairness}, the trust game \citep{berg1995trust}, the public-goods game \citep{samuelson1954pure, andreoni1995cooperation}, the Bomb Risk Elicitation Task \citep{crosetto2013bomb}, the beauty-contest game \citep{nagel1995unraveling,keynes1936general}, experimental Cournot competition \citep{cournot1838recherches,huck1999learning}, and the common-pool-resource game \citep{lloyd1833two,walker1990rent}.

\vspace{3mm}
\begin{promptpart}{situationcol}{Dictator}
You are paired with another player. Your role is to decide how to divide \$100 and the other player simply receives your choice. How would you like to divide the money? Please give only one concrete choice and highlight the amount you give to the other player in [] (such as [\$x]).
\end{promptpart}
\vspace{3mm}

\vspace{3mm}
\begin{promptpart}{situationcol}{Ultimatum game: Proposer role}
This is a two-player game. You are the Proposer, and the other player is the Responder. As the proposer, you propose how to divide \$100 and the Responder chooses either Accept or Reject. If accepted, the two of you will earn as described by the accepted proposal accordingly. If rejected, then both of you will earn \$0. 
How much would you like to propose to give to the Responder? Please give only one concrete choice, and highlight the amount with [] (such as [\$x]).
\end{promptpart}
\vspace{3mm}

\vspace{3mm}
\begin{promptpart}{situationcol}{Ultimatum game: Responder role}
This is a two-player game. You are the Responder, and the other player is the Proposer. The proposer proposes how to divide \$100 and you, as the Responder, choose either Accept or Reject. If accepted, the two of you will earn as described by the accepted proposal accordingly. If rejected, then both of you will earn \$0. 
As the Responder, what is the minimal amount in the proposal that you would accept? Please give only one concrete choice, and highlight the amount with [] (such as [\$x]).
\end{promptpart}
\vspace{3mm}

\vspace{3mm}
\begin{promptpart}{situationcol}{Trust game: Investor role}
This is a two-player game. You are an Investor and the other player is a Banker. You have \$100 to invest and you choose how much of your money to invest with the Banker. The amount you choose to invest will grow by 3x with the Banker. For example, if you invest \$10, it will grow to \$30 with the Banker. The Banker then decides how much of the money (\$0-\$30) to return to you, the Investor.
How much of the \$100 would you like to invest with the Banker? Please give only one concrete choice, and highlight the number with [] (such as [\$x]).
\end{promptpart}
\vspace{3mm}

\begin{promptpart}{situationcol}{Trust game: Banker role}
This is a two-player game. You are a Banker and the other player is an Investor, and the goal for each player is to earn more. The Investor chooses how much of the money (up to \$100) to invest with you. The amount the Investor invests will generate a 2x return with you (the current value of investment becomes 3x). To settle the investment, as the Banker, you get to decide how much of this total amount to return to the Investor and the rest will be kept as your profit. For example, you can choose to return \$0 (therefore the investor will lose their investment), or you can return the entire 3x (initial investment + 2x profit) to the investor, or you can return any amount in between.
Now, if the investor has invested \$50 with you and the current value became \$150, how much of the \$150 would you like to return to the Investor? Please give only one concrete choice, and highlight the number with [] (such as [\$x]).
\end{promptpart}
\vspace{3mm}

\vspace{3mm}
\begin{promptpart}{situationcol}{Public Goods}
In this public good game, you and 3 others will choose how much to contribute towards a water cleaning project. Each of you gets \$20 per round to contribute between \$0 and \$20. The project has a 50\% return rate. Your payoff relies on both your and others' contributions. Everyone benefits from the group's total contribution. Your payoff in each round equals the amount you didn't contribute (endowment - contribution) plus everyone's benefit (total contributions * 50\% return rate). Here are two examples to calculate your payoff.

Example one: You contributed \$12; total group contributions were \$20

Your Payoff = (\$20-\$12) + \$20*50\% = \$18

Example two: You contributed \$12; total group contributions were \$30

Your Payoff = (\$20-\$12) + \$30*50\% = \$23

We will play a total of 3 rounds, in the first round, how much of the \$20 would you like to contribute? Please give a concrete number and highlight it with [] (e.g., [x]).
\end{promptpart}
\vspace{3mm}

\vspace{3mm}
\begin{promptpart}{situationcol}{Bomb}
There are 100 boxes, and one bomb has been randomly placed in 1 of 100 boxes. You can choose to open 0-100 boxes at the same time. If none of the boxes you open has the bomb, you earn points that are equal to the number of boxes you open. If one of the boxes you open has the bomb, you earn zero points. How many boxes would you open? Please give one concrete number and highlight it with [] (such as [x]).
\end{promptpart}
\vspace{3mm}

\vspace{3mm}
\begin{promptpart}{situationcol}{Beauty Contest}
This is a number-guessing game where you will play with a number of other players. Each player chooses an integer between 0 and 100 which is hidden from other players. The player who chooses a number that is closest to 2/3 of the average wins the game. For example, if the average of the guesses is 60 and if you guess 40, i.e. 2/3 of 60, you win the game. What number would you like to guess? Please give a concrete number and highlight it with [] (e.g., [x]).
\end{promptpart}
\vspace{3mm}

\vspace{3mm}
\begin{promptpart}{situationcol}{Cournot}
This is a market game where you and 1 other oil producer compete. Each of you simultaneously chooses how many barrels of oil to produce, from 0 to 15 barrels. You pay a production cost of \$6 for each barrel you produce, and this cost is the same for both producers. The market price per barrel depends on the total production of both producers: the price equals 30 minus the total number of barrels produced. Your profit equals the number of barrels you produce multiplied by (market price - \$6). For example, suppose you produce 10 barrels and the other producer produces 9 barrels, and the resulting market price is \$11 per barrel. Then your profit is 10 * (\$11 - \$6) = \$50, and the other producer's profit is 9 * (\$11 - \$6) = \$45. How many barrels would you like to produce? Please give one concrete number and highlight it with [] (such as [x]).
\end{promptpart}
\vspace{3mm}

\vspace{3mm}
\begin{promptpart}{situationcol}{Commons}
This is a fishing game where you and 4 other players share a common fishing ground. Each of you simultaneously chooses how many hours to fish, from 0 to 48 hours. Your revenue per hour of fishing depends on the total hours fished by the whole group: the more hours fished in total, the lower the revenue per hour. Specifically, the revenue per hour equals (48 * 5) minus the total hours fished by all five players, that is, 240 minus the group's total hours. Your payoff equals the number of hours you fish multiplied by the revenue per hour. For example, if you fish 24 hours and the other four players fish 84 hours in total, the revenue per hour is 240 - 108 = \$132, so your payoff is 24 * \$132 = \$3,168. How many hours would you like to fish? Please give one concrete number and highlight it with [] (such as [x]).
\end{promptpart}
\vspace{3mm}

\subsection{Data Generation and Extraction}
\label{app:provenance}

Table~\ref{tab:provenance} summarizes every model-generated dataset used in the paper.
All responses for the main analyses were generated by \texttt{gpt-4.1}, snapshot \texttt{gpt-4.1-2025-04-14}, through the Azure Batch deployment.
We retained the model's default sampling parameters and did not specify temperature, top-$p$, penalties, or output-token limits.

The model-robustness analysis additionally queried four LLMs: \texttt{gpt-5.6-terra}, \texttt{gpt-5.6-luna}, \texttt{DeepSeek-V4-Pro}, and \texttt{claude-sonnet-5}.
For these models, we disabled reasoning to match \texttt{gpt-4.1} and otherwise retained each provider's default inference settings.
The Claude API required a maximum output length, which we set to 4,096 tokens.
Each model received the same prompts and game instructions as in the corresponding \texttt{gpt-4.1} collections.

\newcommand{\ProvenanceResponseCount}{1{,}258{,}827}
\newcommand{\ProvenanceFallbackCount}{54{,}757}
\newcommand{\ProvenanceFallbackShare}{4.3\%}

\begin{table}[ht]
\centering
\footnotesize
\begin{tabular}{llrl}
\toprule
Dataset & Player model & Responses & Collected \\
\midrule
Default prompt & \texttt{gpt-4.1} & 100 & 2026-03-04 to 2026-08-04 \\
Economic keywords, $k{=}1$--$5$, five-level & \texttt{gpt-4.1} & 777{,}696 & 2026-06-26 to 2026-07-23 \\
Economic keywords, 5-dim grid, three-level & \texttt{gpt-4.1} & 24{,}303 & 2026-07-21 \\
Economic keywords, $k{=}1$--$4$, three-level & \texttt{gpt-4.1} & 78{,}000 & 2026-08-13 \\
Big Five, $k{=}1$--$3$, five-level & \texttt{gpt-4.1} & 152{,}515 & 2026-06-26 to 2026-06-30 \\
Big Five, 5-dim grid, three-level & \texttt{gpt-4.1} & 24{,}300 & 2026-06-09 \\
Big Five, $k{=}1$--$4$, three-level & \texttt{gpt-4.1} & 78{,}000 & 2026-08-13 \\
Placebo keywords, single, five-level & \texttt{gpt-4.1} & 7{,}516 & 2026-08-04 \\
Positive placebo, $k{=}1$--$5$, three-level & \texttt{gpt-4.1} & 102{,}302 & 2026-08-13 \\
Generalizability, four models, five-level & Four models & 14{,}035 & 2026-08-04 to 2026-08-05 \\
\bottomrule
\end{tabular}
\caption{Provenance of the model-generated datasets used in the paper.
All GPT-4.1 rows use snapshot \texttt{gpt-4.1-2025-04-14}.
Response counts include all API replies recorded in each collection, including replacement calls used to complete cells with invalid responses.
Date ranges indicate collections completed in multiple stages.
The generalizability row pools four models: \texttt{gpt-5.6-terra}, \texttt{gpt-5.6-luna}, \texttt{DeepSeek-V4-Pro}, and \texttt{claude-sonnet-5}; its GPT-4.1 comparison reuses the corresponding main-analysis data.}
\label{tab:provenance}
\end{table}

We extracted numeric choices from the raw replies using a fixed pipeline applied to every collection.
A regular expression first identified an unambiguous bracketed numeric choice, which is the output format required in game instructions (Appendix \ref{app:game-instructions}).
Replies that this rule could not resolve were read separately by \texttt{gpt-4.1-nano} and \texttt{gpt-5.6-luna} using the same game-specific extraction prompt.
When the two readers agreed, we accepted their value; every disagreement was adjudicated by hand.
The fallback readers were used for 54{,}757 replies, or 4.3\% of the responses reported in Table~\ref{tab:provenance}.
A reply entered the analytic data only if its API stop signal indicated completion and the extracted choice fell within the game's action range.
Incomplete or out-of-range replies were excluded and replaced.

\newpage \clearpage

\section{Human-playing Data}
\label{app:human-data}

Our human benchmark comes from MobLab, an online platform on which instructors run economic games with their classes.
We use \TotalSubjects who together contributed \TotalDecision across ten game roles.
Each subject contributes at most one choice per game.
The data span 4,875 sessions run between 2014 and 2026.
Sessions are mostly classroom cohorts, so subjects are predominantly university students. 
Table~\ref{tab:human-games} reports the per-game statistics. 

\begin{table}[htbp]
\centering
\small
\setlength{\tabcolsep}{5pt}
\begin{tabular}{lcrrrrrc}
\toprule
Game & Action space & Subjects & Mean & SD & Median & Sessions & Years \\
\midrule
Dictator       & $[0, 100]$ & 10,335 & 25.7 & 21.7 & 25.0 &   722 & 2015--2023 \\
Proposer       & $[0, 100]$ &  5,291 & 44.1 & 20.0 & 49.0 &   189 & 2016--2022 \\
Responder      & $[0, 100]$ &  5,291 & 34.9 & 20.3 & 40.0 &   188 & 2016--2022 \\
Investor       & $[0, 100]$ & 16,830 & 42.0 & 36.1 & 30.0 & 1,207 & 2016--2023 \\
Banker         & $[0, 150]$ &  1,570 & 59.4 & 39.5 & 54.5 &   694 & 2016--2023 \\
Public Goods   & $[0, 20]$  & 18,046 &  8.9 &  5.8 &  9.0 &   701 & 2016--2023 \\
Bomb           & $[0, 100]$ & 23,629 & 45.4 & 24.7 & 49.0 &   620 & 2015--2023 \\
Beauty Contest & $[0, 100]$ & 31,793 & 32.2 & 22.6 & 28.0 & 1,116 & 2015--2023 \\
Cournot        & $[0, 15]$  &  5,023 &  8.5 &  3.2 &  8.0 &   191 & 2023--2026 \\
Commons        & $[0, 48]$  &  1,339 & 20.1 & 14.0 & 20.0 &    77 & 2014--2026 \\
\midrule
All games      &            & 119,147 & \multicolumn{3}{c}{} & 4,875 & 2014--2026 \\
\bottomrule
\end{tabular}
\caption{Statistics of human play data by game. Subject counts are also choice counts: each
subject contributes at most one choice per game.}
\label{tab:human-games}
\end{table}

Table~\ref{tab:human-coverage} gives the distribution of how many games are played by the subjects.
Our individual-level analysis uses subjects who played at least five games, a cohort of \IndivSubjects and who made \IndivDecisions choices. 
No subject played more than eight of the ten games.

\begin{table}[htbp]
\centering
\small
\begin{tabular}{crrrr}
\toprule
& \multicolumn{2}{c}{Exactly $m$ games} & \multicolumn{2}{c}{At least $m$ games} \\
\cmidrule(lr){2-3} \cmidrule(lr){4-5}
Games played $m$ & Subjects & Choices & Subjects & Choices \\
\midrule
1 & 55,022 & 55,022 & 78,657 & 119,147 \\
2 & 14,197 & 28,394 & 23,635 &  64,125 \\
3 &  4,354 & 13,062 &  9,438 &  35,731 \\
4 &  3,350 & 13,400 &  5,084 &  22,669 \\
5 &  1,210 &  6,050 &  1,734 &   9,269 \\
6 &    453 &  2,718 &    524 &   3,219 \\
7 &     67 &    469 &     71 &     501 \\
8 &      4 &     32 &      4 &      32 \\
\bottomrule
\end{tabular}
\caption{Statistics of subjects by number of games played. 
Right-hand columns are
cumulative. Population-level analyses use all \TotalDecision choices; the individual-level analysis uses the $\geq 5$ cohort.}
\label{tab:human-coverage}
\end{table}

\newpage \clearpage

\section{Model Robustness}
\label{app:models}

Throughout the main analysis, we use \texttt{gpt-4.1}.
The method is, however, LLM-agnostic:
The same type of prompts and game instructions can be given to any LLM.
If the dimensions capture general features of human behavior rather than relationships specific to \texttt{gpt-4.1}, we would expect their directional effects to be similar across models.

\begin{figure}[h]
\centering
\includegraphics[width=\linewidth]{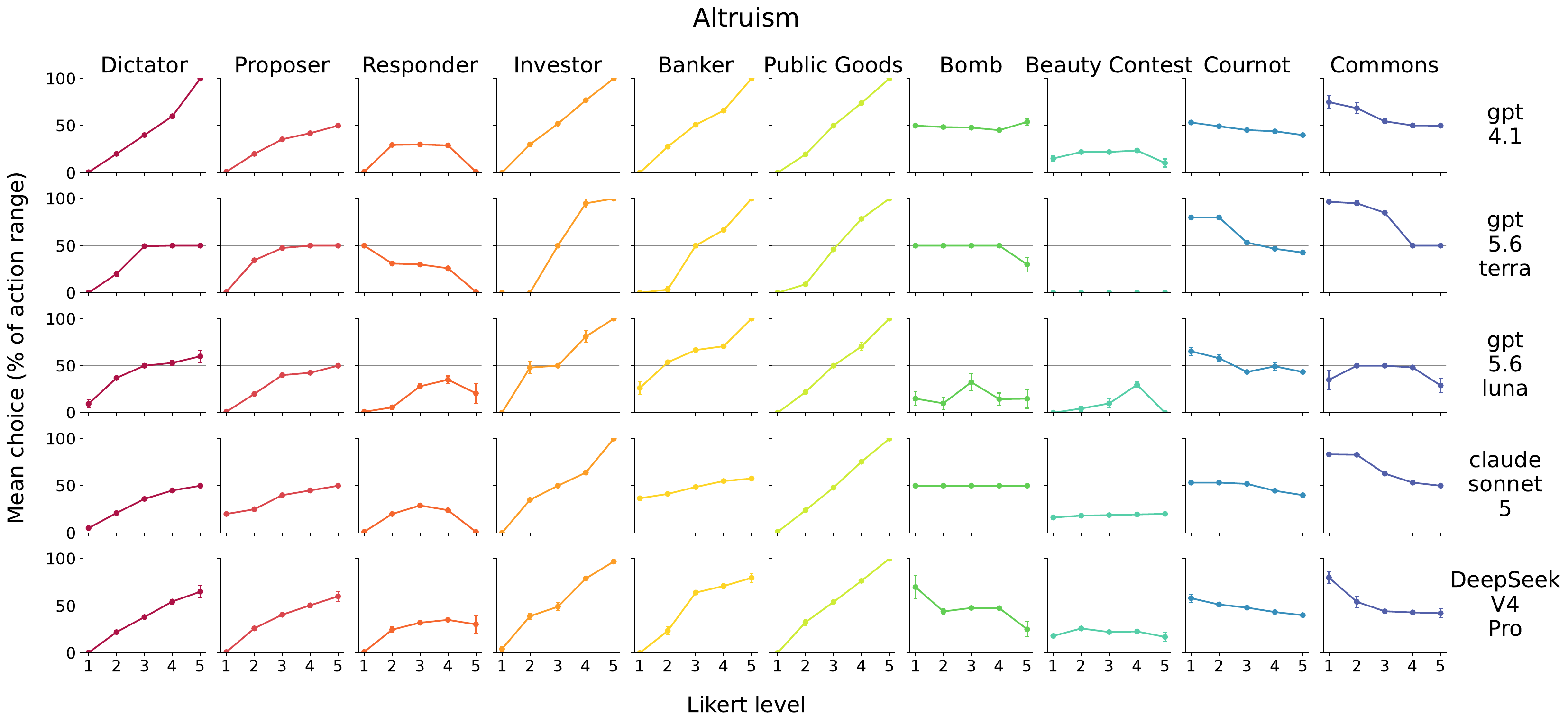}
\caption{This figure reports behavior induced by a single Altruism dimension across LLMs from different developers.
Rows are: Terra, GPT-5.6 Luna, Claude Sonnet 5, and DeepSeek V4 Pro
Columns report the ten game roles.
Points show the mean choice as Altruism varies from 1 to 5, expressed as a percentage of each game’s action range, and error bars show $\pm 1$ standard error across ten responses at each level. 
The gray horizontal line marks the midpoint of the action range.}
\label{fig:dim_model_altruism}
\end{figure}

\begin{figure}
\centering
\includegraphics[width=\linewidth]{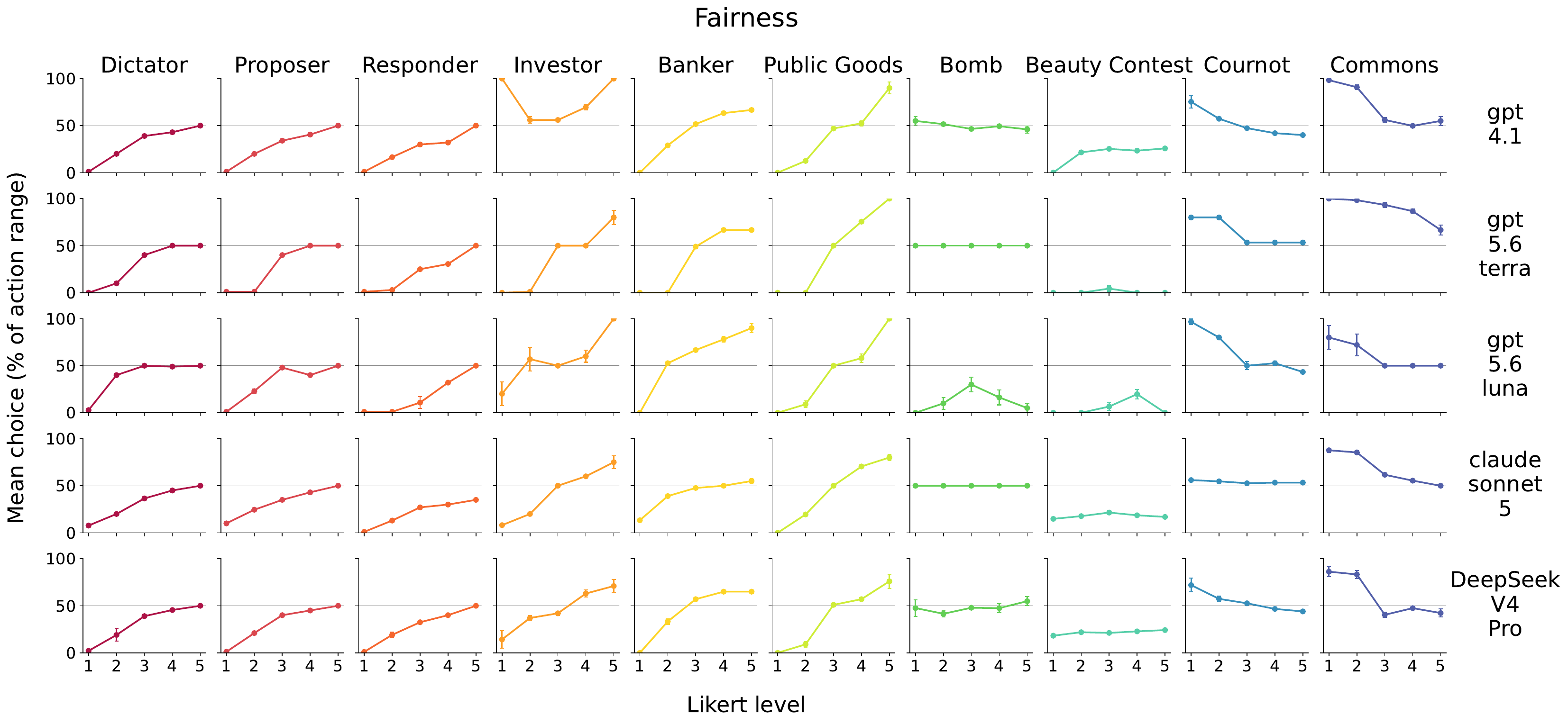}
\caption{This figure reports behavior induced by a single Fairness dimension across LLMs from different developers.
Rows are: Terra, GPT-5.6 Luna, Claude Sonnet 5, and DeepSeek V4 Pro
Columns report the ten game roles.
Points show the mean choice as Fairness varies from 1 to 5, expressed as a percentage of each game’s action range, and error bars show ±1 standard error across ten responses at each level. 
The gray horizontal line marks the midpoint of the action range.}
\label{fig:dim_model_fairness}
\end{figure}

\begin{figure}
\centering
\includegraphics[width=\linewidth]{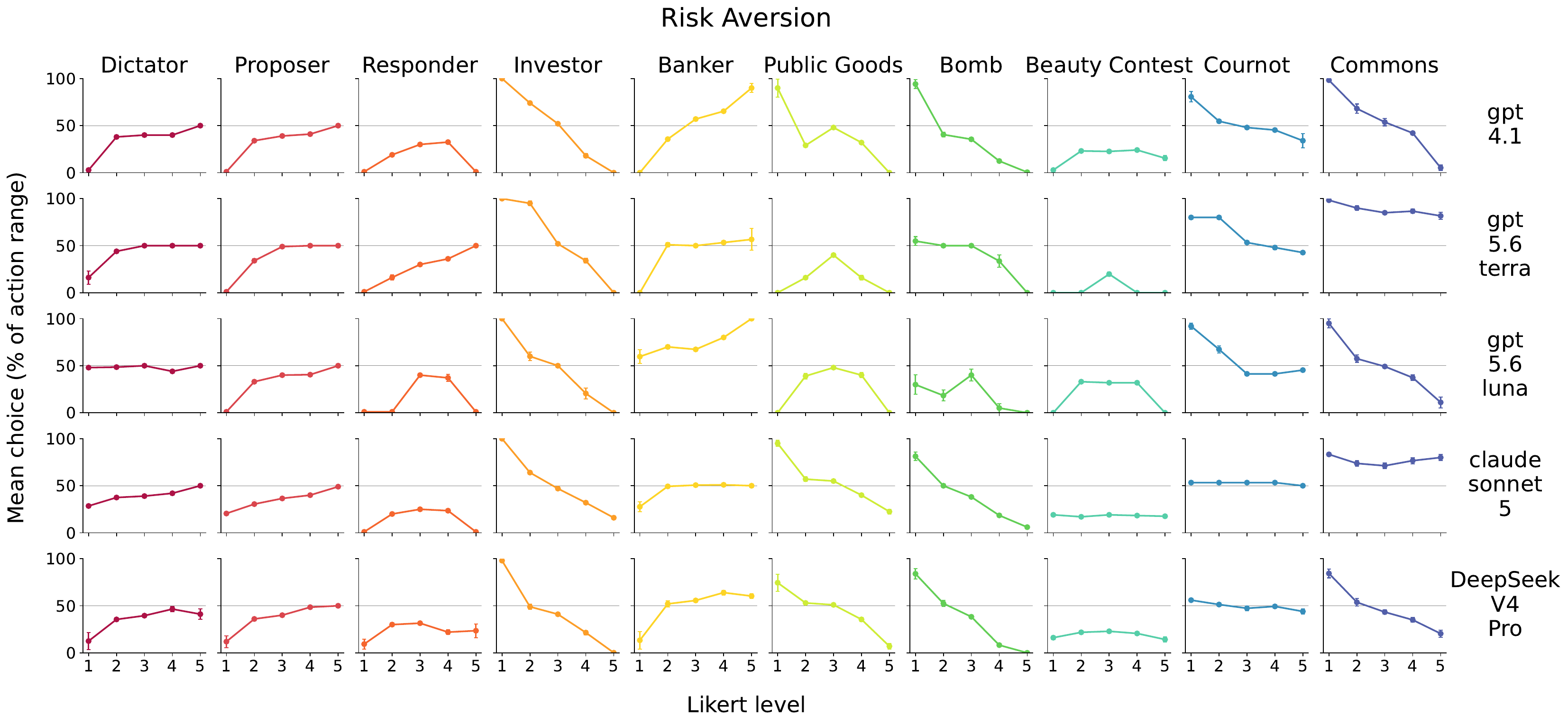}
\caption{This figure reports behavior induced by a single Risk Aversion dimension across LLMs from different developers.
Rows are: Terra, GPT-5.6 Luna, Claude Sonnet 5, and DeepSeek V4 Pro
Columns report the ten game roles.
Points show the mean choice as Risk Aversion varies from 1 to 5, expressed as a percentage of each game’s action range, and error bars show ±1 standard error across ten responses at each level. 
The gray horizontal line marks the midpoint of the action range.}
\label{fig:dim_model_risk}
\end{figure}

\begin{figure}
\centering
\includegraphics[width=\linewidth]{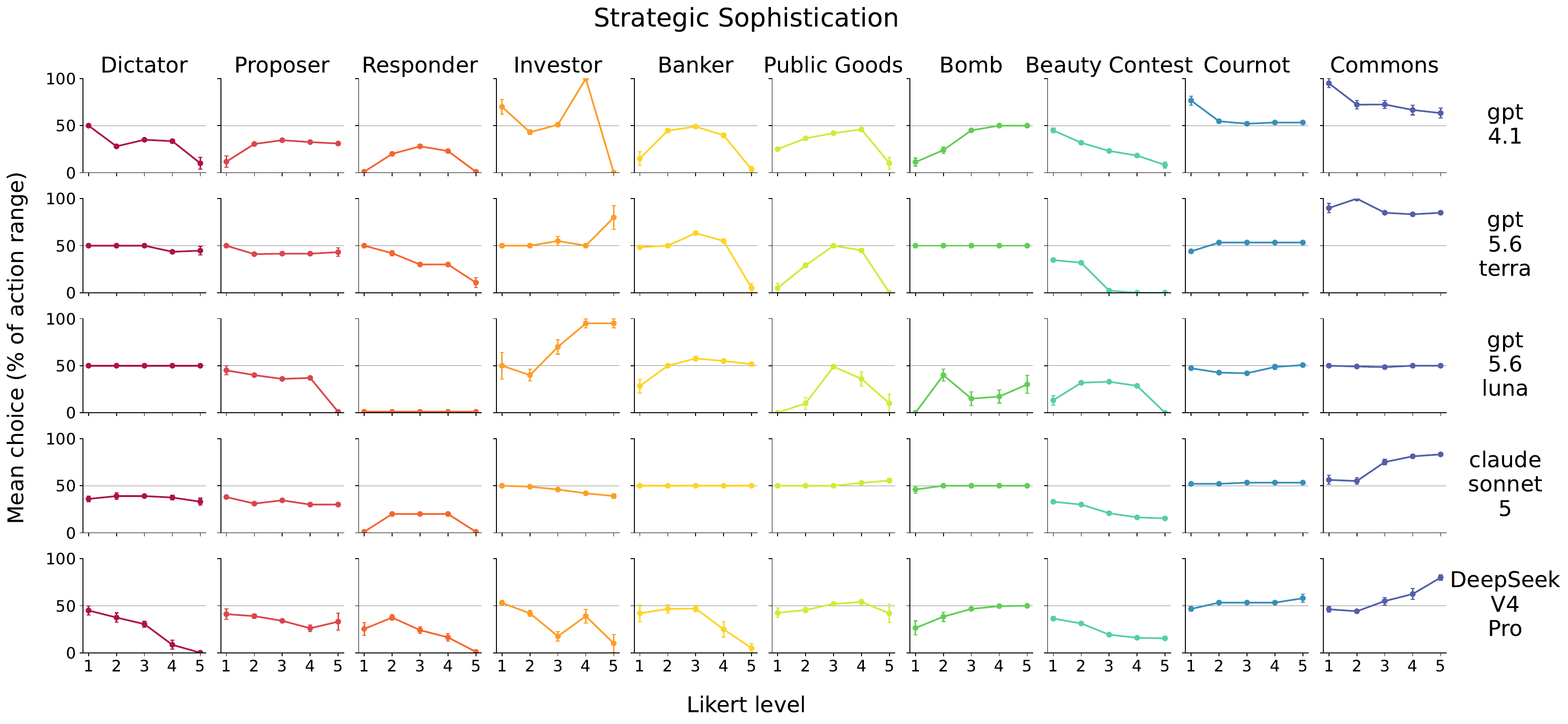}
\caption{This figure reports behavior induced by a single Strategic Sophistication dimension across LLMs from different developers.
Rows are: Terra, GPT-5.6 Luna, Claude Sonnet 5, and DeepSeek V4 Pro
Columns report the ten game roles.
Points show the mean choice as Strategic Sophistication varies from 1 to 5, expressed as a percentage of each game’s action range, and error bars show ±1 standard error across ten responses at each level. 
The gray horizontal line marks the midpoint of the action range.}
\label{fig:dim_model_strategic}
\end{figure}

\begin{figure}
\centering
\includegraphics[width=\linewidth]{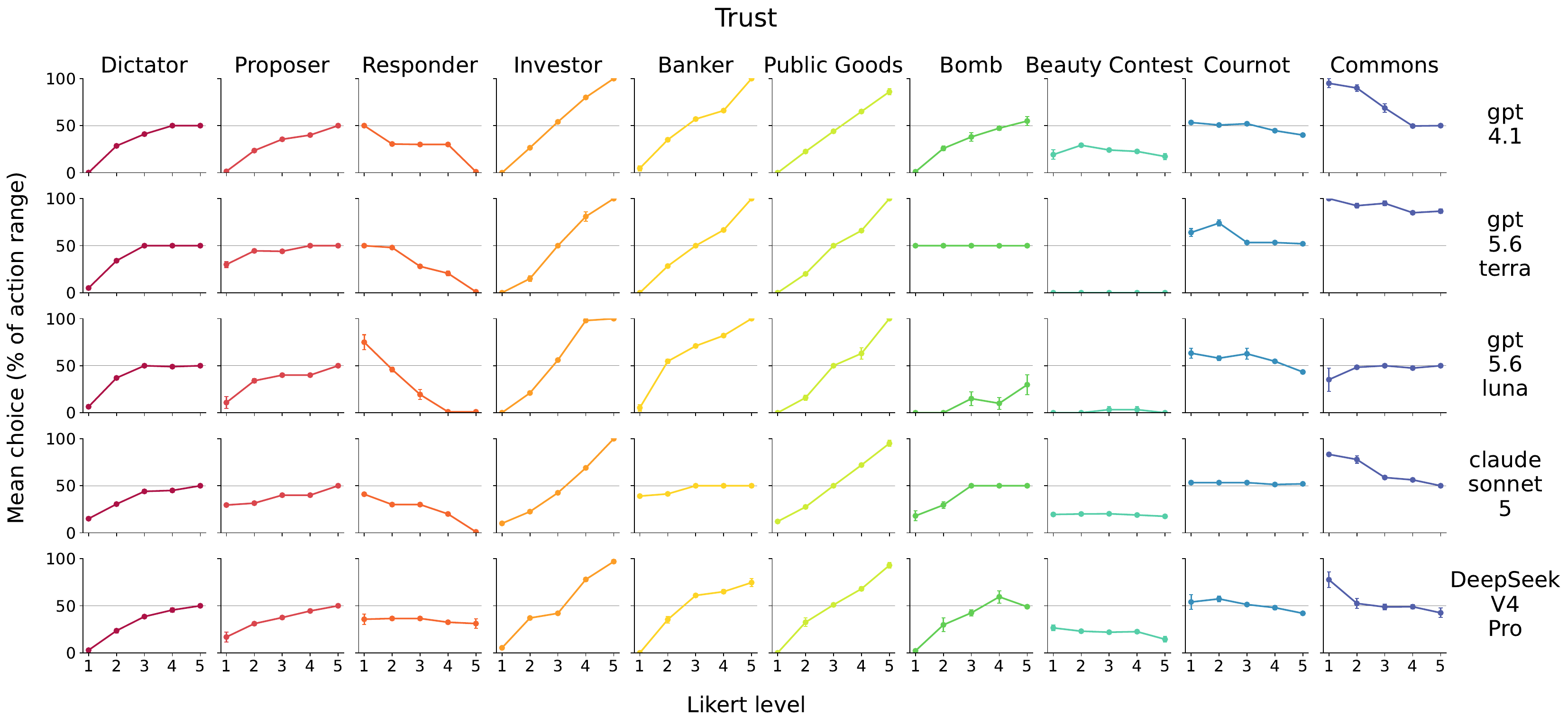}
\caption{This figure reports behavior induced by a single Trust dimension across LLMs from different developers.
Rows are: Terra, GPT-5.6 Luna, Claude Sonnet 5, and DeepSeek V4 Pro
Columns report the ten game roles.
Points show the mean choice as Trust varies from 1 to 5, expressed as a percentage of each game’s action range, and error bars show ±1 standard error across ten responses at each level. 
The gray horizontal line marks the midpoint of the action range.}
\label{fig:dim_model_trust}
\end{figure}

Repeating the full analysis for every LLM would be costly.
We therefore repeat the one-dimensional exercise underlying Figure~\ref{fig:dimension_behavior_L5} using models that differ in capability and developer: \texttt{gpt-4.1}, \texttt{gpt-5.6-terra}, \texttt{gpt-5.6-luna}, \texttt{DeepSeek-V4-Pro}, and \texttt{claude-sonnet-5}.
For each model, we vary each economic characteristic from 1 to 5 and generate ten choices at every level in every game.
Figures~\ref{fig:dim_model_altruism}--\ref{fig:dim_model_trust} show the resulting level-by-level relationships.
We then calculate the Spearman correlation between the prompted level and the resulting choices for each characteristic, game, and model.

\begin{figure}[h]
\centering
\includegraphics[width=\linewidth]{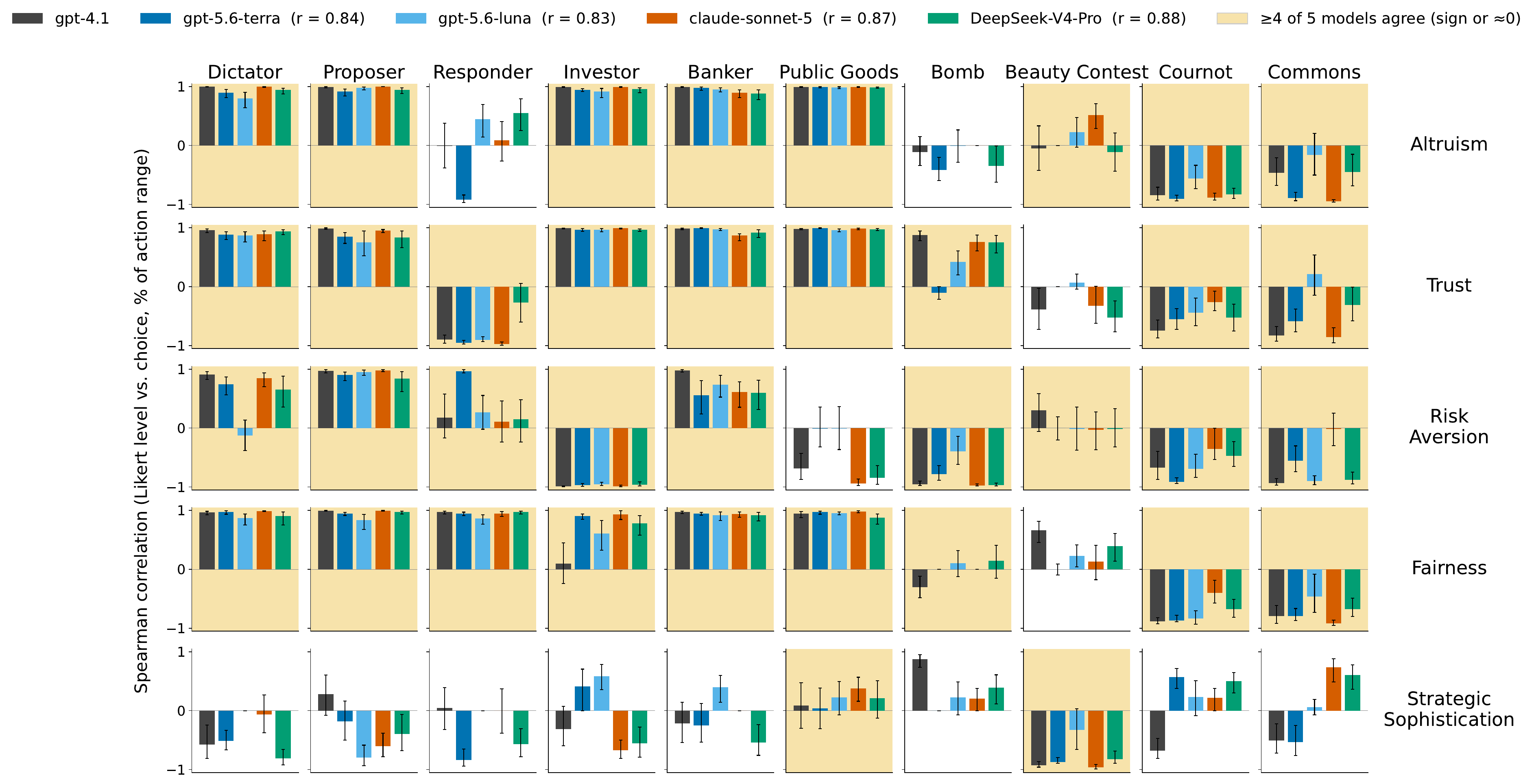}
\caption{This figure reports the effects of the economic traits across LLMs.
Rows report the five economic traits, and columns report the ten game roles.
Within each panel, bars report the Spearman correlation between the trait's Likert level and choices for the LLMs: \texttt{gpt-4.1}, \texttt{gpt-5.6-terra}, \texttt{gpt-5.6-luna}, \texttt{DeepSeek-V4-Pro}, and \texttt{claude-sonnet-5}.
Correlations pool the ten choices generated at each of five Likert levels, for approximately 50 observations per bar, and choices are expressed as a percentage of each game's action range.
Error bars report bootstrap 95\% confidence intervals.
A correlation is reported as zero without an error bar when a model's choices do not vary.
Gold shading indicates that at least four models have significantly positive correlations, significantly negative correlations, or correlations whose confidence intervals include zero.
The $r$ values shown in parentheses beside each alternative model in the legend report correlation-of-correlations between that model's 50 trait-by-game Spearman correlations and the corresponding correlations for \texttt{gpt-4.1}.}
\label{fig:lfive-correlation}
\end{figure}

Figure~\ref{fig:lfive-correlation} summarizes these correlations across models.
The gold shading identifies panels in which at least four models fall into the same category: significantly positive, significantly negative, or not statistically distinguishable from zero.
The $r$ values shown beside the alternative models in the legend report the Pearson correlation between that model's 50 trait-by-game Spearman correlations and the corresponding correlations for \texttt{gpt-4.1}.

The dimensions have similar effects across the five models.
The correlation-of-correlations for the alternative models compared to \texttt{gpt-4.1} are high, ranging from $r=0.83$ to $r=0.88$.
Furthermore, in 37 of the 50 trait-by-game comparisons, at least four models have significantly positive correlations, significantly negative correlations, or correlations that are not statistically distinguishable from zero.
For Altruism, Trust, Risk Aversion, and Fairness, this is true in 35 of the 40 comparisons.
The agreement is especially clear where the characteristics have large and intuitive effects.
For example, higher Altruism increases giving or investment in the Dictator, Investor, Banker, and Public Goods games across all five models, while having little effect in the Bomb game.
Together, these comparisons show agreement both panel by panel and in the overall pattern of effects across characteristics and games.

Strategic Sophistication is the main exception, with mostly unshaded panels.
However, in the Beauty Contest, where it should matter most, its effect is strongly negative and consistent across models.
And elsewhere its relationships tend to be weaker than many of those observed in other panels.

\newpage \clearpage

\section{Behavior induced by two-dimensional type vectors}
\label{app:bivariate-effects}

The figures below examine every two-characteristic combination among the five economic characteristics.
Each figure treats one characteristic as the focal characteristic, which varies from 1 to 5 along the $x$-axis.
The columns report the ten game roles, and the rows report the other four characteristics.
Within each row, the colored lines hold the second characteristic fixed at each of its five levels.
The dashed gray line shows behavior when the focal characteristic appears alone in a one-dimensional type vector.

\begin{figure}[h]
\centering
\includegraphics[width=\linewidth]{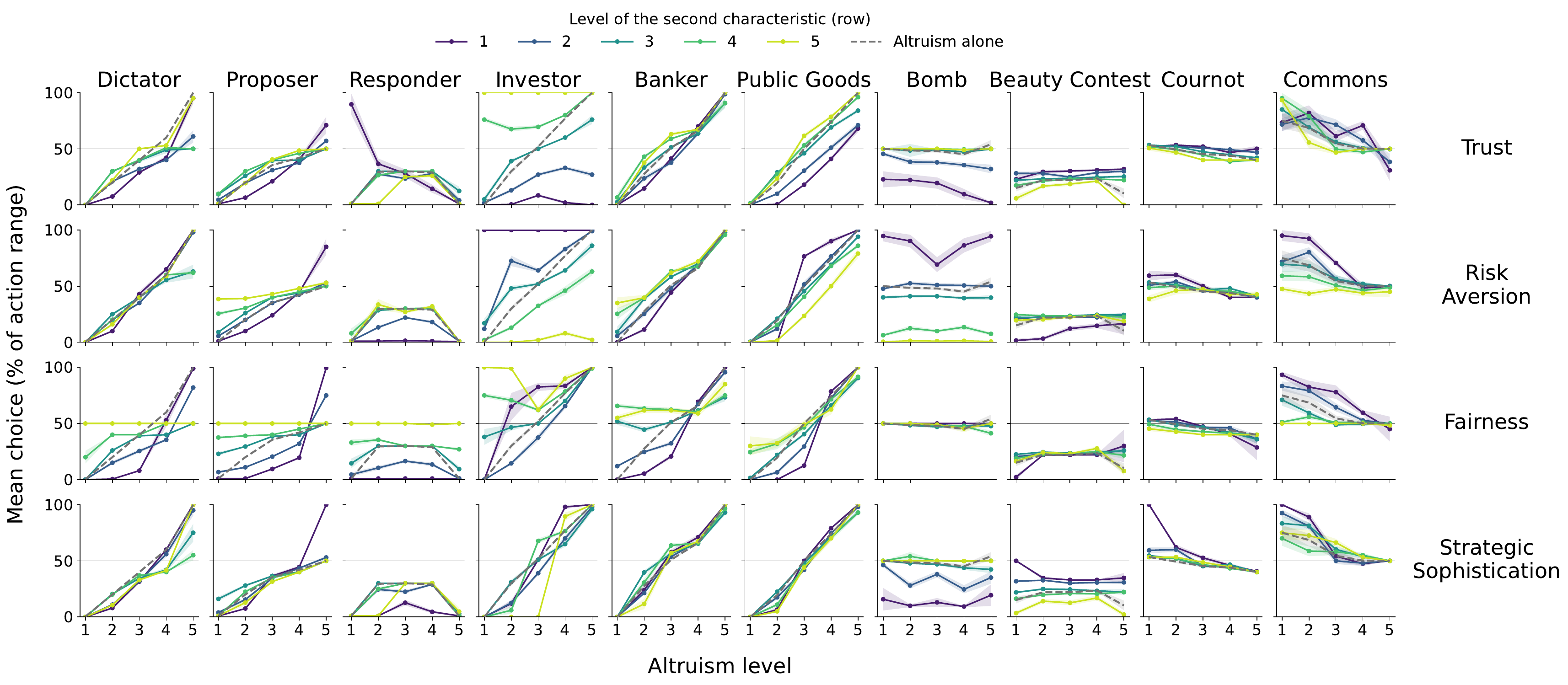}
\caption{This figure shows behavior as Altruism varies within two-dimensional type vectors.
The $x$-axis varies Altruism from 1 to 5.
Each row pairs Altruism with the second characteristic named at left, and the colored lines hold the level of that characteristic fixed from 1 to 5.
The dashed gray line shows behavior when Altruism appears alone in a one-dimensional type vector.
Each column reports a game role.
Points show the mean choice as a percentage of the game's action range, and shaded bands show $\pm 1$ standard error across ten responses.}
\label{fig:bivariate-altruism}
\end{figure}

\begin{figure}[h]
\centering
\includegraphics[width=\linewidth]{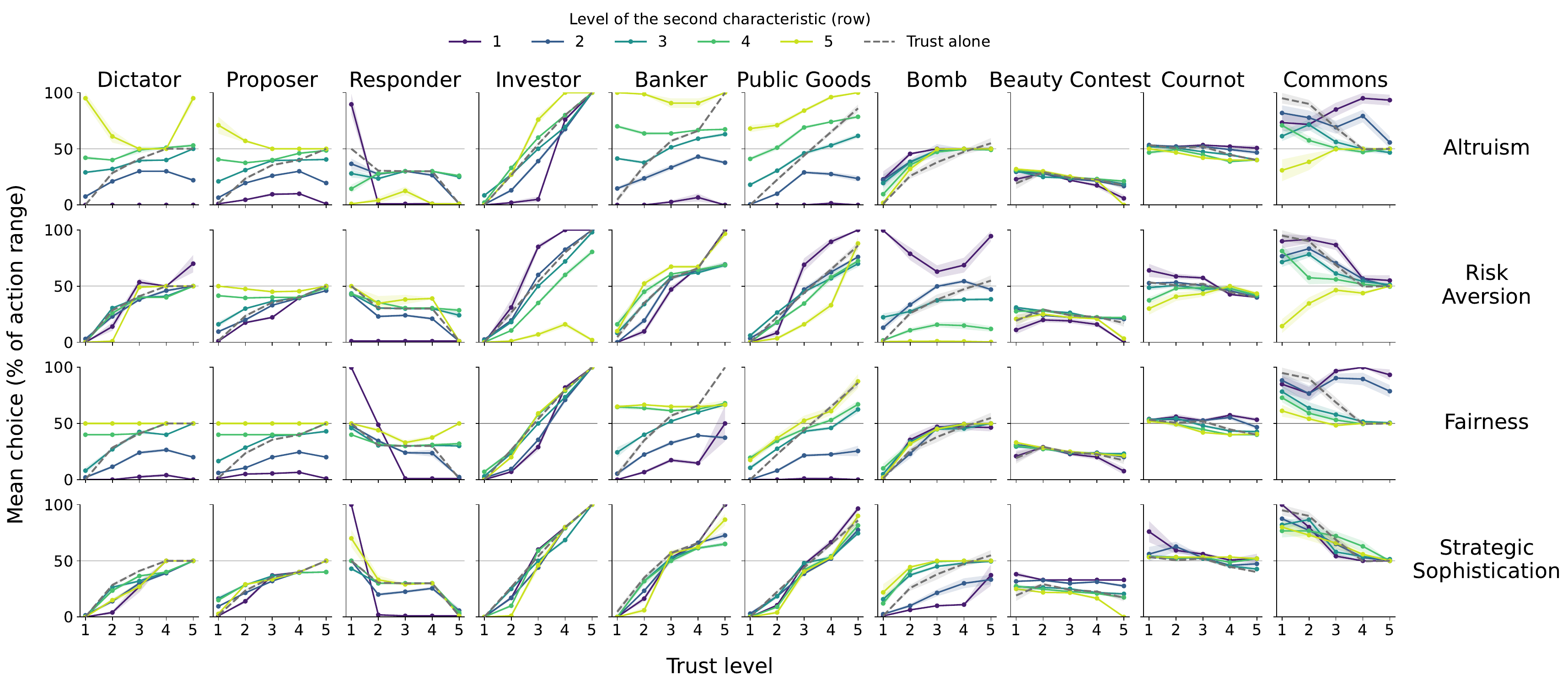}
\caption{This figure shows behavior as Trust varies within two-dimensional type vectors.
The $x$-axis varies Trust from 1 to 5.
Each row pairs Trust with the second characteristic named at left, and the colored lines hold the level of that characteristic fixed from 1 to 5.
The dashed gray line shows behavior when Trust appears alone in a one-dimensional type vector.
Each column reports a game role.
Points show the mean choice as a percentage of the game's action range, and shaded bands show $\pm 1$ standard error across ten responses.}
\label{fig:bivariate-trust}
\end{figure}

\begin{figure}[p]
\centering
\includegraphics[width=\linewidth]{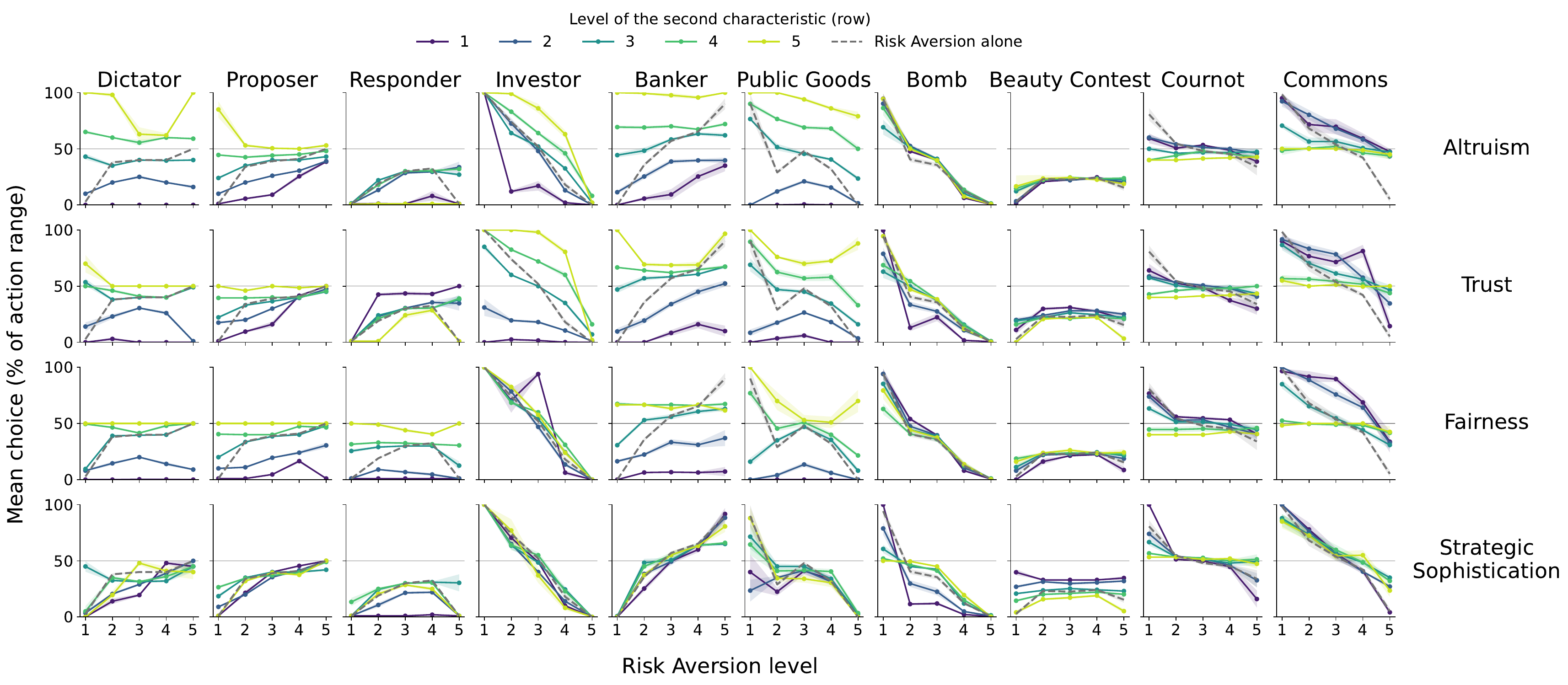}
\caption{This figure shows behavior as Risk Aversion varies within two-dimensional type vectors.
The $x$-axis varies Risk Aversion from 1 to 5.
Each row pairs Risk Aversion with the second characteristic named at left, and the colored lines hold the level of that characteristic fixed from 1 to 5.
The dashed gray line shows behavior when Risk Aversion appears alone in a one-dimensional type vector.
Each column reports a game role.
Points show the mean choice as a percentage of the game's action range, and shaded bands show $\pm 1$ standard error across ten responses.}
\label{fig:bivariate-risk-aversion}
\end{figure}

\begin{figure}[p]
\centering
\includegraphics[width=\linewidth]{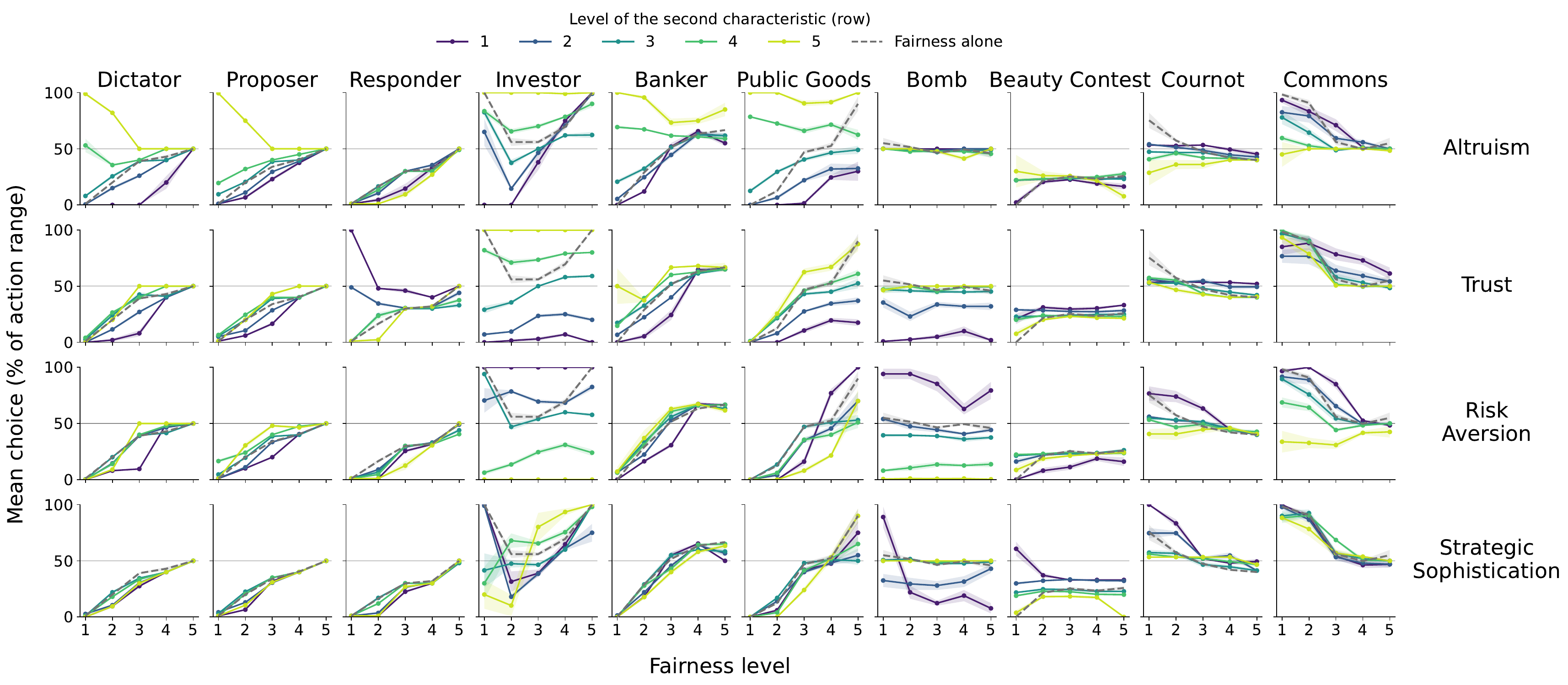}
\caption{This figure shows behavior as Fairness varies within two-dimensional type vectors.
The $x$-axis varies Fairness from 1 to 5.
Each row pairs Fairness with the second characteristic named at left, and the colored lines hold the level of that characteristic fixed from 1 to 5.
The dashed gray line shows behavior when Fairness appears alone in a one-dimensional type vector.
Each column reports a game role.
Points show the mean choice as a percentage of the game's action range, and shaded bands show $\pm 1$ standard error across ten responses.}
\label{fig:bivariate-fairness}
\end{figure}

\begin{figure}[p]
\centering
\includegraphics[width=\linewidth]{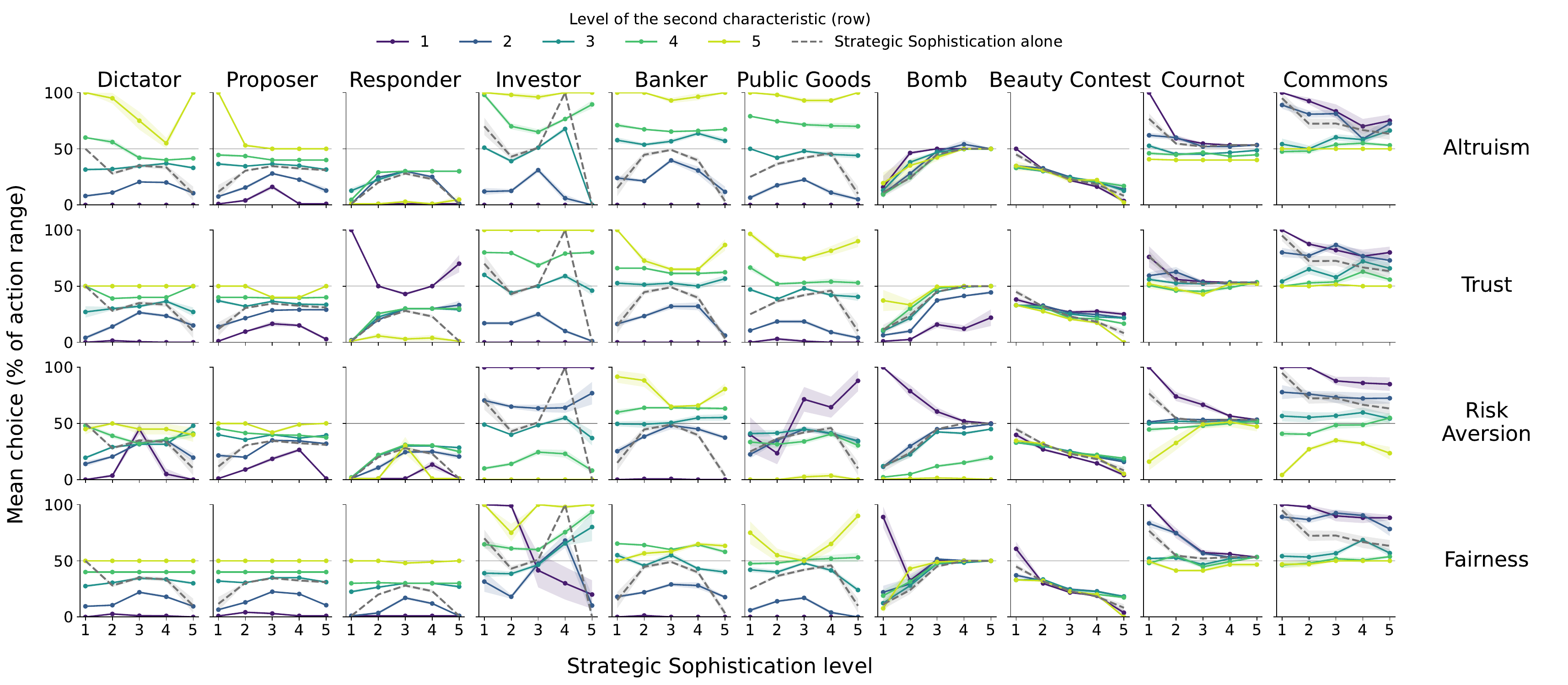}
\caption{This figure shows behavior as Strategic Sophistication varies within two-dimensional type vectors.
The $x$-axis varies Strategic Sophistication from 1 to 5.
Each row pairs Strategic Sophistication with the second characteristic named at left, and the colored lines hold the level of that characteristic fixed from 1 to 5.
The dashed gray line shows behavior when Strategic Sophistication appears alone in a one-dimensional type vector.
Each column reports a game role.
Points show the mean choice as a percentage of the game's action range, and shaded bands show $\pm 1$ standard error across ten responses.}
\label{fig:bivariate-strategic-sophistication}
\end{figure}

\newpage \clearpage

\section{Additional Figures}
\label{app:figs}

\begin{figure}[h]
\centering
\includegraphics[width=\linewidth]{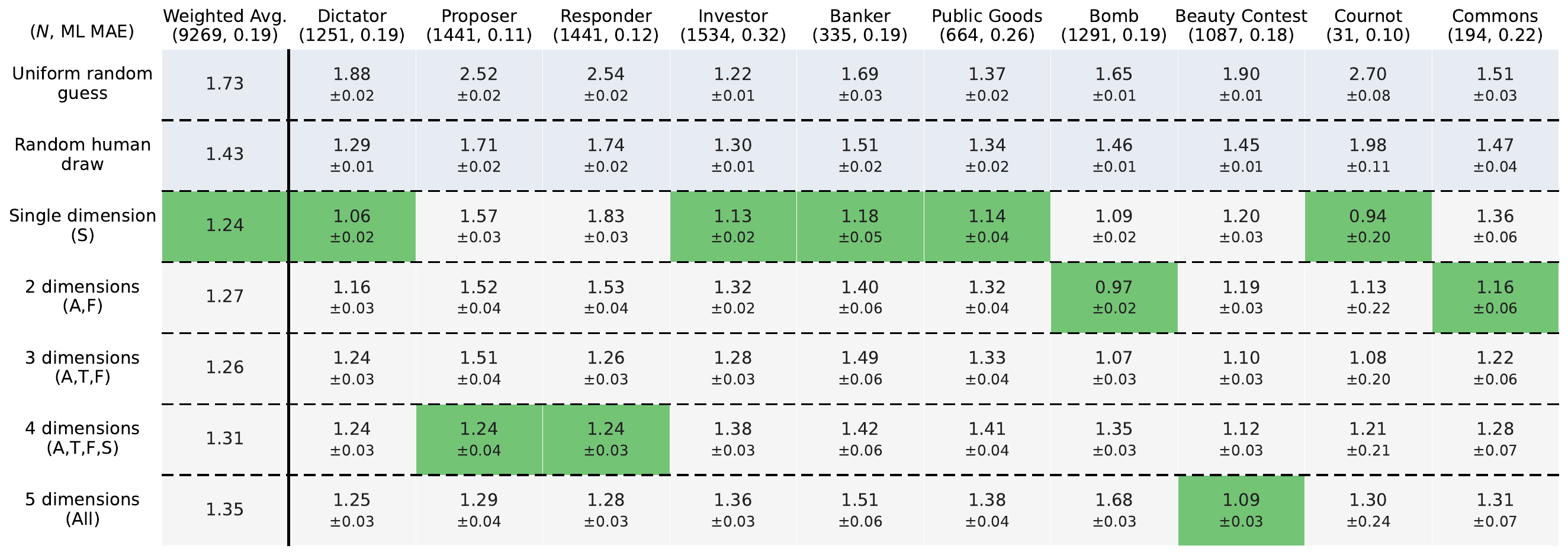}
\caption{Held-out-game prediction error relative to HistGBT.
Each row reports, for a given number of dimensions, the combination with the lowest sample-weighted mean held-out error across games.
These combinations are selected using the same held-out results summarized in the figure.
For each subject and held-out game, the subject's type is estimated without using their choice in that game.
The type-based prediction is the median of the ten choices generated by the matched type in the held-out game.
When several types fit equally well, their held-out prediction errors are averaged.
HistGBT predicts the held-out choice from the subject's choices in the other games and is evaluated using five-fold cross-validation across subjects.
Each cell reports the mean type-based error divided by the mean HistGBT error.
All errors are mean absolute errors measured as a share of the game's action range.
Column headers report $(N, \text{ML MAE})$: the number of subjects who played the game and the mean HistGBT error, which is the denominator of every ratio in that column.
Note that for the Weighted average, $N$ is the number of decisions.
The uniform-random row reports the expected error from predicting a uniformly random value on the game's action range.
The random-human row reports the expected error from using a randomly selected human's choice as the prediction.
The highlighted cell identifies the lowest relative error among the five type rows in each column.
Numbers following $\pm$ report the standard error of the corresponding mean error divided by the HistGBT mean error.}
\label{fig:fit-hist-ratio}
\end{figure}

\begin{figure}[h]
\centering
\includegraphics[width=.8\linewidth]{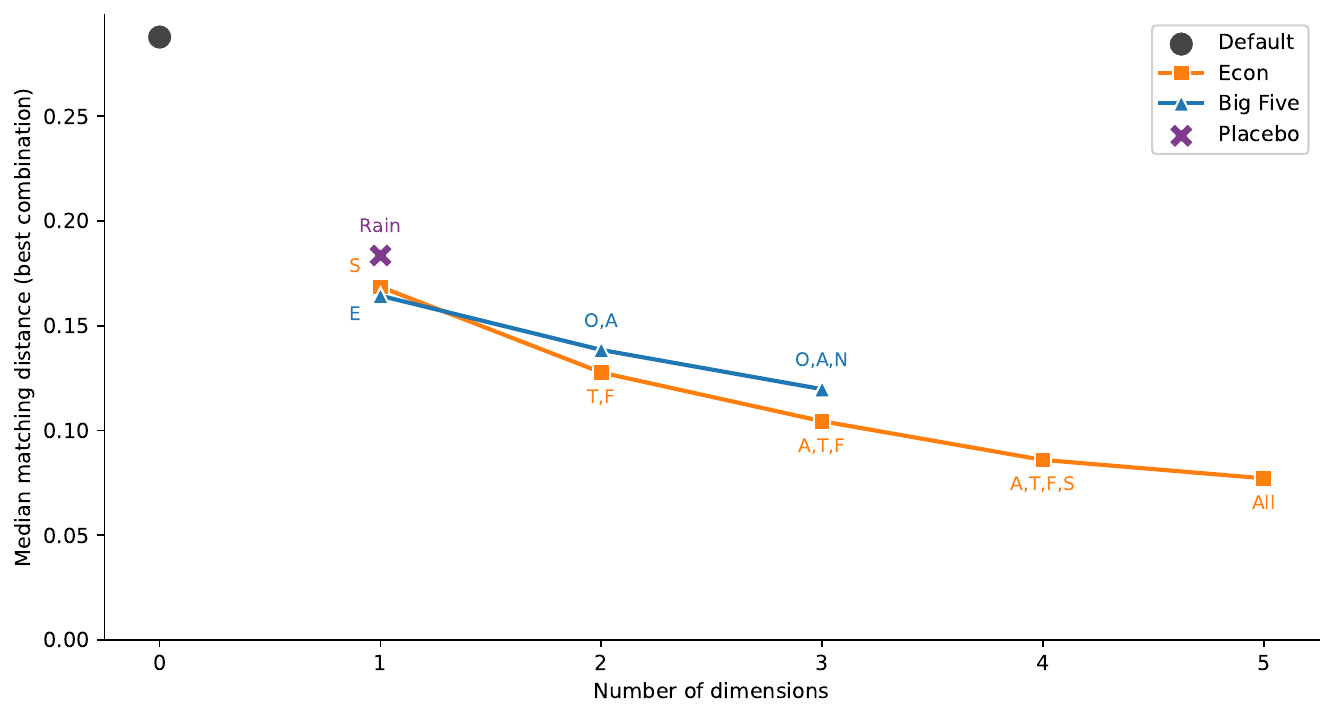}
\caption{The comparison of Figure~\ref{fig:alt-key-L3} repeated on a 5-level Likert scale. Median matching distance across subjects by number of dimensions, for the five economic characteristics, the Big-5 OCEAN personality traits, and the placebo traits, all on a 5-level Likert scale. 
For each number of dimensions $k$, each point reports the combination of $k$ characteristics from that set with the lowest median matching distance across subjects, where each subject is matched by their best-fitting type vector within that combination. The label beside each point names the winning combination by the initial letters of its characteristics.
The figure is incomplete because of limited funds to trace out the full curves for the placebo and Big 5.
}
\label{fig:alt-key-L5}
\end{figure}

\begin{figure}[h]
\centering
\includegraphics[width=\linewidth]{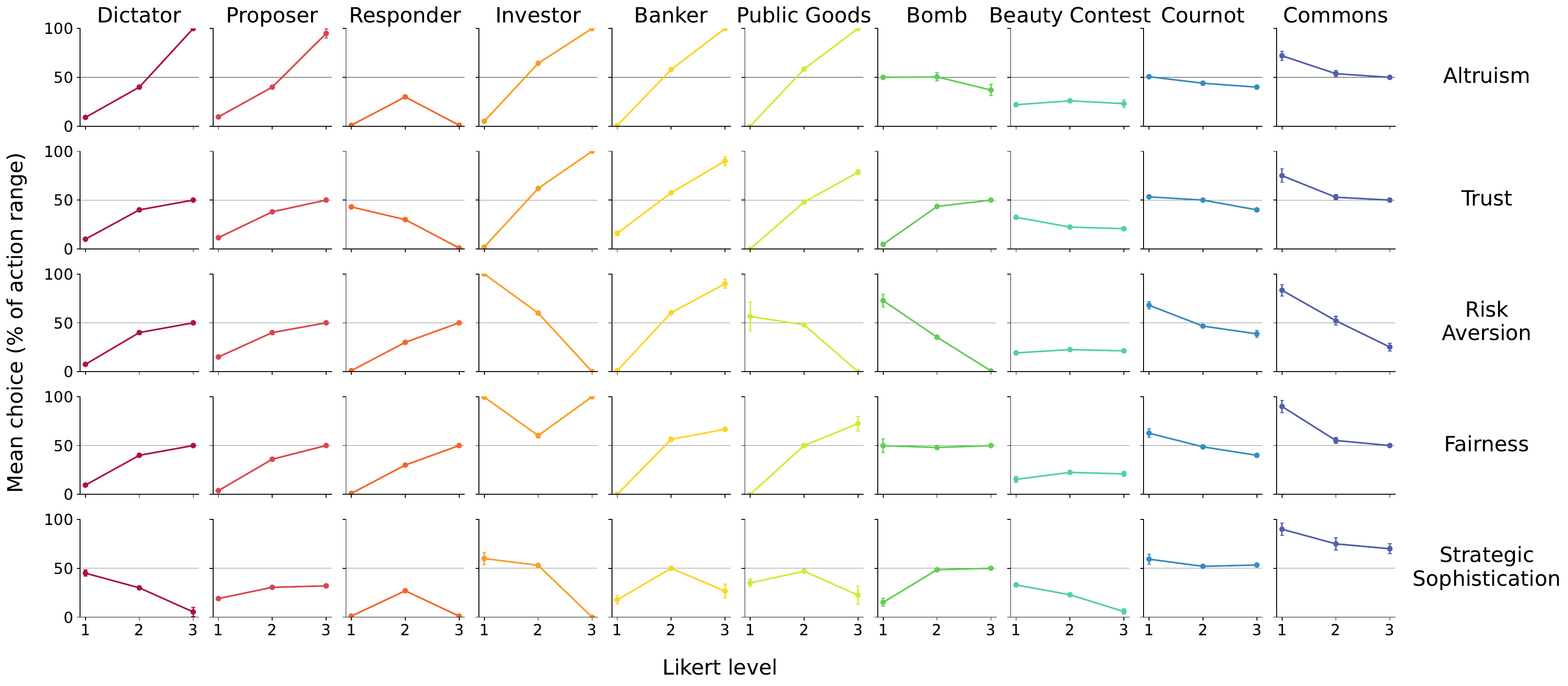}
\caption{This figure reports behavior induced by the five economic key words on a 3-level Likert scale. Each row varies one placebo trait from 1 to 3, and each column reports a game role. 
Points show the mean choice as a percentage of the game’s action range, and error bars show $\pm$1 standard error across ten responses
at each level.
The gray horizontal line marks the midpoint of the action range.}
\label{fig:dim_econ_Lthree}
\end{figure}

\begin{figure}[h]
\centering
\includegraphics[width=\linewidth]{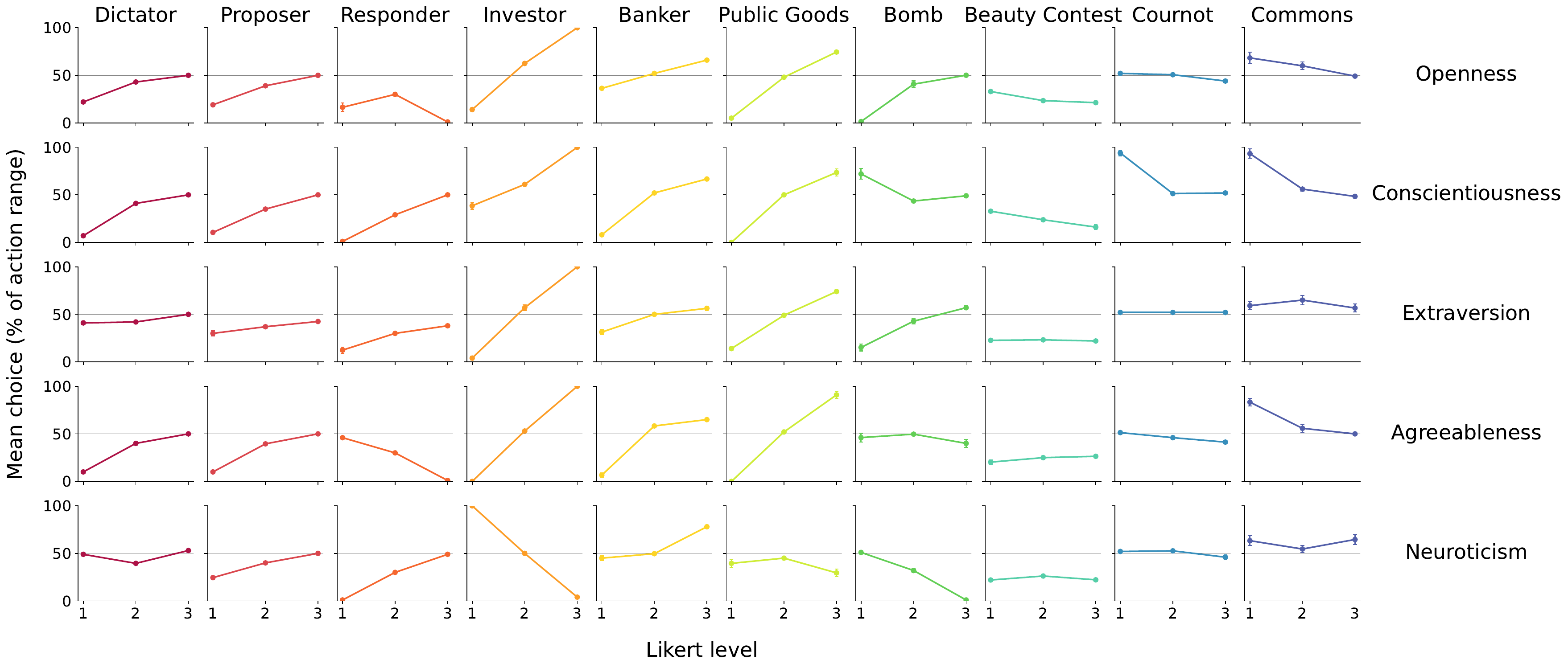}
\caption{This figure reports behavior induced by the big 5 psychology key words on a 3-level Likert scale. Each row varies one placebo trait from 1 to 3, and each column reports a game role. 
Points show the mean choice as a percentage of the game’s action range, and error bars show $\pm$1 standard error across ten responses
at each level.
The gray horizontal line marks the midpoint of the action range.}
\label{fig:dim_big5_Lthree}
\end{figure}

\begin{figure}[h]
\centering
\includegraphics[width=\linewidth]{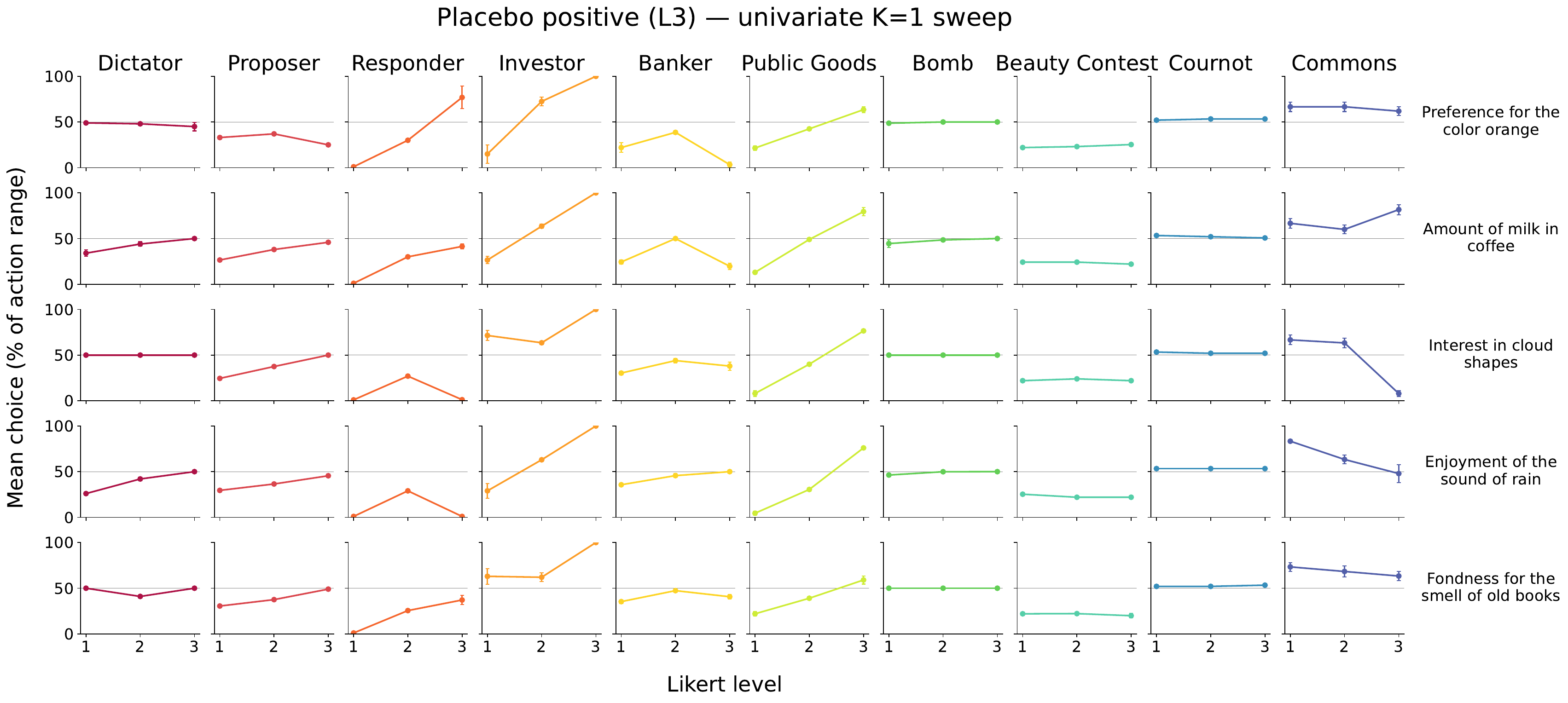}
\caption{This figure reports Behavior induced by placebo traits. Each row varies one placebo trait from 1 to 3, and each column reports a game role. 
Points show the mean choice as a percentage of the game’s action range, and error bars show $\pm$1 standard error across ten responses
at each level.
The gray horizontal line marks the midpoint of the action range.}
\label{fig:dim_placebo_Lthree}
\end{figure}

\end{document}